\documentclass[twocolappendix,numberedappendix]{emulateapj}
\usepackage{amsmath}
\usepackage{float}   % Allows me to specify where I want my figures to go
\usepackage{color}
\usepackage{rotating}
\usepackage{placeins} % I have an inordinate amount of floats that causes Latex to break.  making a \FloatBarrier is the only way I can think to solve this issue
\usepackage{comment}
\usepackage{epsfig,graphicx}
\usepackage{epstopdf}

\def\etal{{et~al.}}
\def\kms{{\hbox{km s$^{-1}$}}}

\shortauthors{Grasha \etal}
\slugcomment{Accepted for Publication in the Astrophysical Journal}

\begin{document}
\title{The Hierarchical Distribution of the Young Stellar Clusters in Six Local Star Forming Galaxies}
\author{K. Grasha\altaffilmark{1}, 
D. Calzetti\altaffilmark{1}, 
A. Adamo\altaffilmark{2}, 
H. Kim\altaffilmark{3},  
B.G. Elmegreen\altaffilmark{4}, 
D.A. Gouliermis\altaffilmark{5,6}, 
D.A. Dale\altaffilmark{7}, 
M. Fumagalli\altaffilmark{8}, 
E.K. Grebel\altaffilmark{9}, 
K.E. Johnson\altaffilmark{10}, 
L. Kahre\altaffilmark{11},
R.C. Kennicutt\altaffilmark{12}, 
M. Messa\altaffilmark{2}, 
A. Pellerin\altaffilmark{13}, 
J.E. Ryon\altaffilmark{14}, 
L.J. Smith\altaffilmark{15}, 
F. Shabani\altaffilmark{8}, 
D. Thilker\altaffilmark{16},
L. Ubeda\altaffilmark{14}
}
\altaffiltext{1}{Astronomy Department, University of Massachusetts, Amherst, MA 01003, USA; kgrasha@astro.umass.edu}
\altaffiltext{2}{Dept. of Astronomy, The Oskar Klein Centre, Stockholm University, Stockholm, Sweden}
\altaffiltext{3}{Gemini Observatory, La Serena, Chile}

\altaffiltext{4}{IBM Research Division, T.J. Watson Research Center, Yorktown Hts., NY}
\altaffiltext{5}{Zentrum f\"ur Astronomie der Universit\"at Heidelberg, Institut f\"ur Theoretische Astrophysik, Albert-Ueberle-Str.\,2, 69120 Heidelberg, Germany}
\altaffiltext{6}{Max Planck Institute for Astronomy,  K\"{o}nigstuhl\,17, 69117 Heidelberg, Germany}
\altaffiltext{7}{Dept. of Physics and Astronomy, University of Wyoming, Laramie, WY}
\altaffiltext{8}{Institute for Computational Cosmology and Centre for Extragalactic Astronomy, Durham University, Durham, United Kingdom}
\altaffiltext{9}{Astronomisches Rechen-Institut, Zentrum f\"ur Astronomie der Universit\"at Heidelberg, M\"onchhofstr.\ 12--14, 69120 Heidelberg, Germany}
\altaffiltext{10}{Dept. of Astronomy, University of Virginia, Charlottesville, VA}
\altaffiltext{11}{Dept. of Astronomy, New Mexico State University, Las Cruces, NM}
\altaffiltext{12}{Institute of Astronomy, University of Cambridge, Cambridge, United Kingdom}
\altaffiltext{13}{Dept. of Physics and Astronomy, State University of New York at Geneseo, Geneseo NY}
\altaffiltext{14}{Space Telescope Science Institute, Baltimore, MD}
\altaffiltext{15}{European Space Agency/Space Telescope Science Institute, Baltimore, MD}
\altaffiltext{16}{Dept. of Physics and Astronomy, The Johns Hopkins University, Baltimore, MD}

\begin{abstract}
We present a study of the hierarchical clustering of the young stellar clusters in six local (3--15 Mpc) star-forming galaxies using Hubble Space Telescope broad band WFC3/UVIS UV and optical images from the Treasury Program LEGUS (Legacy ExtraGalactic UV Survey).  We have identified 3685 likely clusters and associations, each visually classified by their morphology, and we use the angular two-point correlation function to study the clustering of these stellar systems.  We find that the spatial distribution of the young clusters and associations are clustered with respect to each other, forming large, unbound hierarchical star-forming complexes that are in general very young.  The strength of the clustering decreases with increasing age of the star clusters and stellar associations, becoming more homogeneously distributed after $\sim$40--60 Myr and on scales larger than a few hundred parsecs.  In all galaxies, the associations exhibit a global behavior that is distinct and more strongly correlated from compact clusters.  Thus, populations of clusters are more evolved than associations in terms of their spatial distribution, traveling significantly from their birth site within a few tens of Myr whereas associations show evidence of disruption occurring very quickly after their formation.  The clustering of the stellar systems resembles that of a turbulent interstellar medium that drives the star formation process, correlating the components in unbound star-forming complexes in a hierarchical manner, dispersing shortly after formation, suggestive of a single, continuous mode of star formation across all galaxies.  
%The clustering properties and the dispersal timescales of the unbound structures suggest a single, continuous mode of star formation.  
\end{abstract}

%It is intriguing to see the similarities of the clustering behavior across such a wide range of galactic stellar masses.

\section{Introduction}\label{sec:intro}
Star clusters are gravitationally bound stellar structures, with radii between 0.5 to several parsecs and masses between $10^3$ and $10^7$~$M_{\odot}$ \citep{portegieszwart10}.  Because most, if not all, stars form in some type of stellar aggregate \citep{lada03}, stellar clusters are a direct product of the star formation process within galaxies.  Compounded by the fact that young stellar clusters are intrinsically brighter than single stars, star clusters become important tracers of the recent star formation history in galaxies beyond which individual stars cannot be detected.  %Furthermore, stellar aggregates and clusters are ``clustered'' with respect to each over, forming extensive star-forming complexes over a large dynamical range that relate to each other in a hierarchical manner. 

Within the hierarchical model, star formation occurs within structures that have smoothly varying densities and sizes that range from pc to kpc scales, with denser regions nested within larger, less dense areas \citep[e.g., ][]{elmegreen06,bastian07}.  Bound star clusters form at these peak densities within the hierarchy.  Most structures within the hierarchy are themselves gravitationally unbound and the stellar components are expected to inherit their clustered substructure from the molecular clouds from which they are born \citep{scalo85}.  Recent analyses of 12 local galaxies \citep{elmegreen14} found that the clustering of star formation remains scale-free, up to the largest scales observable, for both starburst galaxies and more quiescent star-forming galaxies.  This result is consistent within the framework where the self-similar structure of the interstellar medium (ISM), regulated by turbulence, is believed to be the primary driver for the hierarchical nature of star formation \citep{elmegreen96,elmegreen14}.  Thus, extensive star-forming regions of several hundred parsecs or larger are expected to represent common structures, related in both space and time in a hierarchical manner that determines the structure and morphology of all galaxies.  

The evolution and erasure of the unbound hierarchical structures has been the focus of investigation in recent years, where observations of local galaxies support an age-dependent clustering of the stellar components \citep[e.g., ][]{pellerin07,bastian09,scheepmaker09,gieles11,pellerin12,baumgardt13,gouliermis14,gouliermis15,grasha15}, and the clustering becomes progressively weaker for older populations.  Hierarchical clustering is expected to dissipate with age \citep{elmegreen06,elmegreen10} as the densest regions with the shortest mixing timescales lose their substructures first, whereas the larger, unbound regions will lose their substructure over longer periods of time owing to tidal forces and random velocities \citep{bate98}, dispersing over time to form the field population.  Characterizing the clustered nature of star formation provides insight into how star formation is organized across a galaxy, by correlating local environmental conditions at sub-galactic scales -- such as feedback and turbulence -- to the global properties -- such as dynamics and morphology -- of entire galaxies and constrain the migration timescale for which stars and clusters abandon their natal structure.  This will in-turn provide a vital connection between the inherently different processes of clustered star formation seen within local galaxies and the large kpc-scale star-forming structures that appear to be common at high-redshift \citep{immeli04,elmegreen09,forsterschreiber11,guo12}.  

In this paper, we study the young stellar cluster populations of six galaxies as part of the Legacy ExtraGalactic UV Survey\footnote{https://legus.stsci.edu/} \citep[LEGUS;][]{calzetti15}, a Cycle 21 Hubble Space Telescope (HST) program with images of 50 nearby ($\sim$3.5--15~Mpc) galaxies in five UV and optical bands (NUV,U,B,V,I) with the UV/Visible (UVIS) channel on the Wide Field Camera 3 (WFC3) and re-using archival ACS images when appropriate.  The aim of LEGUS is to investigate the relation between star formation and its galactic environment in nearby galaxies, over scales ranging from individual star systems to kpc-sized structures.  These data will help to establish a more accurate picture of galaxy formation and the physical underpinning of the gas-star formation relation.  The relatively nearby location of these galaxies provides us with the high-angular resolution needed to acquire large numbers of star clusters to perform statistically accurate tests for changes in clustering strength across a representative range of galactic environments.  Investigations of a few galaxies from the LEGUS project have already observationally demonstrated the relatively young ($\sim$40--60 Myr) dispersal timescales of star-forming structures \citep{gouliermis15,grasha15}.  

This work builds on our previous paper \citep{grasha15} on a study of the nearby star-forming galaxy NGC 628 using the two-point correlation function as a tool to quantify the clustering properties of the young stellar clusters, finding that the youngest clusters are spatially clustered within unbound, star-forming complexes that disperse with time.  In this work, we expand our sample to investigate the clustering distribution of the stellar clusters within a larger sample of galaxies and a wider range of galactic environments.  We will use the correlation function to identify common age structures, the extent that the distribution of clusters is hierarchical, on which timescale it disperses, and the dependencies of global properties (galaxy type) has on the clustering results, if any.  This will in turn inform on the nature of local, resolved star formation.  % We will investigate and the effect of the local environment -- turbulence -- in a future paper.  

The galaxy selection is described in Section \ref{sec:sample} and the cluster identification process is described in Section \ref{sec:clusterselection}.  The methodology of two-point correlation function is introduced in Section \ref{sec:2pcf}.  In Section \ref{sec:results}, we describe the results and analysis and how we use the correlation function to draw conclusions about the properties of our star clusters.  We discuss our results concerning hierarchy of the stellar clusters in Section \ref{sec:discussion}.  Finally, we summarize the findings of this study in Section \ref{sec:summary}.

\section{Sample Selection}\label{sec:sample}
\begin{figure*}
%\epsscale{1.1}
\includegraphics[scale=.9,bb=0 0 549 651]{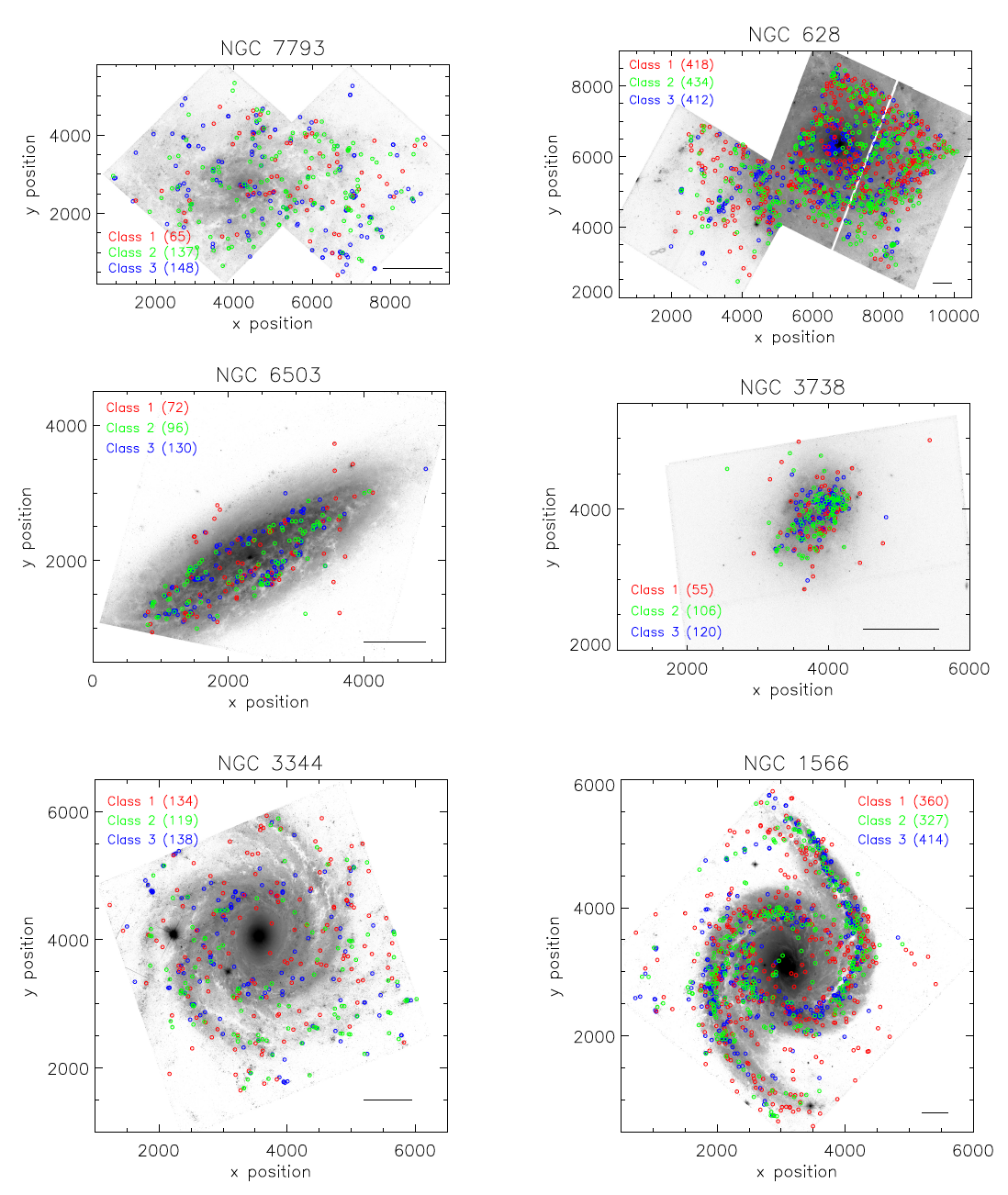}
\caption{
V-band images of each of the six galaxies overlaid with the positions of star cluster and association candidates, where the color corresponds to the morphological classification as described in Section \ref{sec:clusterselection}.  The numbers in parenthesis represent the total number of clusters within the classification. The solid black line in the bottom right of each figure represents the spatial scale of 1 kpc at the distance of that galaxy. 
\label{fig:gal}}
\end{figure*}

In this paper, we select six local ($<$13 Mpc) galaxies, ranging from dwarf to grand design spirals, with visually identified stellar cluster catalogs available (see Section \ref{sec:clusterselection}), from the LEGUS survey.  The galaxies and their general properties are listed in Table \ref{tab1} and shown in Figure \ref{fig:gal} along with the clusters.  All galaxies were observed in five broad band filters: NUV, U, B, V, and I; the list of filters used can be found in \citet{calzetti15}.  Both NGC 628 and NGC 7793 have two pointings, combined into a single mosaic for analysis; the remaining galaxies have one pointing.  %All observations for the NUV and U band were observed with the WFC3 F275W and WFC3 F336W filters, respectively.  

\begin{deluxetable*}{lcccccccccc}
\tablecolumns{11}
%\rotate
\tabletypesize{\scriptsize}
\tablecaption{Properties of the LEGUS Galaxies Sample \label{tab1}}
\tablewidth{0pt}
\tablehead{
\colhead{Name} 	& 
\colhead{Morph.}& 
\colhead{T} 		& 
\colhead{Inclin.} 	& 
\colhead{Dist.} 	& 
\colhead{SFR(UV)}& 
\colhead{M$_*$} & 
\colhead{$\Sigma_{\rm SFR}$} 	&
\colhead{R$_{25}$} &
\colhead{Scale}	&
\colhead{CI$_{\rm cut}$}	
\\
\colhead{} 					& 
\colhead{} 					& 
\colhead{} 					& 
\colhead{(deg.)} 	& 
\colhead{(Mpc)} 			& 
\colhead{(M$_{\odot}$~yr$^{-1}$)} & 
\colhead{(M$_{\odot}$)} & 
\colhead{(M$_{\odot}$ yr$^{-1}$ kpc$^{-2}$)} &
\colhead{(arcm.)} &
\colhead{(pc arcs.$^{-1}$)} &
\colhead{(mag)} 					
\\
\colhead{(1)} & 
\colhead{(2)} & 
\colhead{(3)} & 
\colhead{(4)} &
\colhead{(5)} & 
\colhead{(6)} & 
\colhead{(7)} & 
\colhead{(8)} &
\colhead{(9)} &
\colhead{(10)} &
\colhead{(11)} 
}
\startdata
NGC 7793 & SAd		& 7.4(0.6) & 47.4 & 3.44 	& 0.52 	& 3.2$\times10^{9}$ 	& 0.00907 & 4.67 & 16.678 & 1.3(e)/1.4(w) \\ 
NGC 3738 & Im			& 9.8(0.7) & 40.5 & 4.90 	& 0.07 	& 2.4$\times10^{8}$ 	& 0.01187 & 1.26 & 23.756 & 1.4 \\
NGC 6503 & SAcd		& 5.8(0.5) & 70.2 & 5.27	& 0.32 	& 1.9$\times10^{9}$ 	& 0.00620 & 3.54 & 25.550 & 1.25 \\
NGC 3344 & SABbc	& 4.0(0.3) & 23.7 & 7.0 		& 0.86 	& 5.0$\times10^{9}$ 	& 0.00558 & 3.54 & 33.937 & 1.35 \\
NGC 628   & SAc		& 5.2(0.5) & 25.2 & 9.9 	 	& 3.67 	& 1.1$\times10^{10}$	& 0.00444 & 5.23 & 47.956 & 1.4(c)/1.3(e) \\
NGC 1566 & SABbc	& 4.0(0.2) & 37.3 & 13.2   	& 5.67	& 2.7$\times10^{10}$	& 0.01285 & 4.16 & 64.995 & 1.35
\enddata
\tablecomments{
Columns list the 
(1) galaxy name (ordered in increasing distance); 
(2) morphological type as listed in NED, the NASA Extragalactic Database; 
(3) RC3 morphological T--type as listed in Hyperleda (http://leda.univ-lyon1.fr); 
(4) inclination, in degrees; 
(5) redshift-independent distance in Mpc, 
(6) star formation rate (M$_{\odot}$~yr$^{-1}$), calculated from the GALEX far-UV, corrected for dust attenuation as described in \citet{lee09}; 
(7) stellar masses (M$_{\odot}$) obtained from the extinction-corrected B-band luminosity and color information as described in \citet{bothwell09};  
(8) star formation rate surface density $\Sigma_{\rm SFR}$; %, calculated over the field of view for each galaxy with correction for internal attenuation applied using the method from \citet{hao11}; 
(9) optical radius of the galaxy R$_{25}$ from \citet{devaucouleurs91} in arcmin;  
(10) scale in pc arcsec$^{-1}$; and
(11) concentration index (CI) cutoff between stars and star clusters (see Section \ref{sec:clusterselection}).  Both NGC 7793 and NGC 628 were observed with two pointings and the different CI cutoff for each pointing [central (c), east (e), or west (w)] is given. 
\tablenotemark
}
\end{deluxetable*}

%NGC 7793e	& 	 1.3 \\
%NGC 7793w&  1.4 \\
%NGC 3738	&  1.4\\
%NGC 6503	& 	1.25 \\
%NGC 3344	& 1.35 \\
%NGC 628c 	& 1.4 \\
%NGC 628e	& 	1.3 \\
%NGC 1566	& 	1.35 

%NGC 1313 & SBd		& 7.0(0.4) & 40.7 & 4.39 	& 1.15 		& 2.6$\times10^{9}$ & 2.1$\times10^{9}$ \\
%NGC 5253 & Im			& 11 		 & 67.7 & 3.15  	& 0.10 		& 2.2$\times10^{8}$ & 1.0$\times10^{8}$ \\

%7793:  0.66
%3738: 0.94
%6503: 1.1
%3344: 1.3
%628:  1.9
%1566: 2.5

\subsection{NGC 7793}
NGC 7793 is a spiral galaxy in the Sculptor group at a distance of 3.44 Mpc classified as morphological type SAd.  With an angular size of 9\farcm3 x 6\farcm3, both a west and east pointing were observed to cover a significant portion of the galaxy.  A study of the resolved stars has found that the radial profile exhibits a break at 5.1~kpc; beyond the disk break, the younger populations exhibit a steeper profile, indicative of high levels of stellar radial migration \citep{radburnsmith12}.  %At the distance of NGC 7793, one pixel corresponds to a resolution of 0.66 parsecs.  

\subsection{NGC 3738}
NGC 3738 is an irregular dwarf galaxy that makes up part of the Messier 81 group, located at 4.9 Mpc.  NGC 3738 is a very small system with an apparent size of 2\farcm6 x 2\farcm2 and is the only Irregular galaxy in this study. %giving a resolution of 0.94 parsecs per pixel

\subsection{NGC 6503}
NGC 6503 is a ring dwarf spiral galaxy of morphological type SAcd located at a distance of 5.27 Mpc.  The region of intense star formation in the galaxy is sufficiently compact to be fully covered with a single pointing and is observed to be organized into a ring \citep{knapen06}.  Work by \citet{gouliermis15} shows that younger stars are organized in a hierarchical distribution whereas older stars display a homogeneous, less clustered distribution with a structure dispersion timescale of $\sim60$ Myr.  %, giving a resolution of 1.1 parsecs per pixel

\subsection{NGC 3344}
NGC 3344 is an isolated barred spiral galaxy with a morphological classification of SABbc, located at a distance of 7 Mpc.  With apparent dimensions of 7\farcm1 x 6\farcm5, we have a single pointing of NGC 3344 that covers most of the inner region of the galaxy.  There is a presence of ring-like morphological features at 1 kpc and 7 kpc, with a small bar present within the inner ring \citep{verdes-montenegro00}.  % with an image resolution of 1.3 parsec per pixel

\subsection{NGC 628}
NGC 628 is a morphological type SAc face-on grand design spiral galaxy located at a distance of 9.9~Mpc with no apparent bulge.  The largest galaxy of the M74 galaxy group with an angular size of 10\farcm5 x 9\farcm5, a central and east pointing were obtained.  The clustering of the young stellar clusters in this galaxy was investigated in an earlier paper by \citet{grasha15}, where an age dependency for the clustering strength was found with a randomization time scale of $\sim40$~Myr, after which the clusters displayed a flatter, more homogenous distribution.  %, with an HST resolution of 1.9 pc per pixel

\subsection{NGC 1566}
The most distant galaxy in our sample and the brightest member of the Dorado group, NGC 1566 is a face-on spiral galaxy with an intermediate-strength bar type classified as a SABbc.  NGC 1566 is shown to host a low-luminosity AGN \citep{combes14} with a star forming ring at 1.7 kpc \citep{smajic15}.  We acknowledge that the distance of NGC 1566 is uncertain in the literature; in this work, we adopt the value of 13.2 Mpc, as reported in \citet{calzetti15}.  %The corresponding resolution is 2.5 parsecs per pixel
%has the highest global SFR(UV) in our sample of 5.67 $M_{\odot}~{\rm yr}^{-1}$ and 

%NGC 1313 & SBd		& 7.0(0.4) & 40.7 & 4.39 	& 1.15 		& 2.6E09 & 2.1E09 \\
%NGC 5253 & Im			& 11 		 & 67.7 & 3.15  	& 0.10 		& 2.2E08 & 1.0E08 \\

\section{Cluster Identification, Selection, and Characterization}\label{sec:clusterselection}
A general description of the standard data reduction of the LEGUS datasets is available in \citet{calzetti15}.  A detailed description of the cluster selection, identification, photometry and SED fitting for the LEGUS galaxies is presented in \citet{adamo17}.  Here we summarize the aspects of that paper that are relevant for the current analysis. 

Stellar clusters within each galaxy are identified first through an automated process using SExtractor \citep{bertin96} from the white-light image produced with the five standard LEGUS bands (see \citet{calzetti15} for the method used to produce white-light images).  The configurations of SExtractor are optimized to detect a minimum of a $4\sigma$ source in at least 5 contiguous pixels within a region 30x30 pixels in size.  These initial catalogs are visually inspected to see if any obvious clusters are excluded, improving on the initial SExtractor parameters.  

The photometry within each band, for all sources, is corrected for foreground Galactic extinction \citep{schlafly11}.  With the exception of NGC 7793, photometry is performed with a circular aperture of 4 pixels (3.8 to 10 parsecs) in radius, with the background measured within an annulus of 7 pixels (7--18 parsecs) in inner radius and 1 pixel in width.  NGC 7793 uses an aperture radius of 5 pixels (3.3 parsec) since it is our nearest galaxy.  Each automatic catalog includes sources which satisfy the two following conditions: (1) the V band concentration index (CI; difference in magnitudes measured with an aperture of radius 1 pixel compared to that within 3 pixels) must be greater than the stellar CI peak value; and (2) the source must be detected in at least two contiguous filters (the reference V band and either B or I band) with a photometric error $\sigma_{\lambda} \leq 0.35$ mag.  The first condition above is intended to minimize stellar contamination from the automatic generated cluster catalog.  The second condition corresponds to cluster candidate with a signal-to-noise greater than 3; this is imposed to obtain reliable constraints on the derived cluster properties (age, mass, and extinction).  The median errors on the photometry of the star clusters in NGC 628 are in the range 0.04--0.08 mag.

As stars are unresolved even at the highest HST resolution power, their CI will vary little and their distribution will be highly peaked around an average value typical of a stellar PSF within the same galaxy.  Additionally, stellar clusters are partially resolved and their sizes can vary, therefore, on average, they will have larger CI values than stellar values.  Using CI distributions of the extracted cluster candidates, we select a CI value that separates cluster candidates from the bulk of the stellar interlopers within each system.  As the resolution power is dependent on both the HST camera used and the distance of the galaxy, the values used for the CI cutoff will vary from galaxy to galaxy and for different HST cameras.  Table \ref{tab1} lists the cutoff CI values between stars and clusters for each frame.  

The LEGUS automatic cluster catalog is further down-selected based on the availability of SED fitting and visual classification outcome.  In order to secure reliable measurements from SED fitting, each source is required to have a detection in at least four of the five photometric bands, necessary to adequately break the age-extinction degeneracy.  The physical properties of each cluster (age, extinction, and mass) are derived using deterministic stellar population models \citep[Yggdrasil;][]{zackrisson11} and a $\chi^2$ fitting approach which includes uncertainty estimates \citep[see ][]{adamo10}.  The uncertainties of the age and mass estimates have errors of $\sim$0.1 dex.  To analyze the cluster populations, we use models with solar metallicity for both stars and gas, an average covering factor of 50\%, and a starburst attenuation curve.  The Appendix \ref{sec:appA} details how the adopted attenuation curve affects the derived ages for the star clusters, however, the adoption of different curves only minimally impacts the clustering results.  

Within each galaxy, we visually inspect the subsample of the automatic cluster candidates that have an absolute magnitude brighter than $-6$ mag in the V band (excluding NGC 1566, where the detection limit is $-8$ mag).  The magnitude limit is introduced according to the detection limits of the LEGUS sample, which enables selecting down to $\sim$1000 M$_{\odot}$, 6 Myr old clusters with color excess E(B--V)= 0.25 \citep{calzetti15}.  The magnitude cut is imposed on the aperture-corrected F555W magnitudes for all the LEGUS galaxies, with the exception of NGC 1566, which we applied an absolute V magnitude of $-8$, required to reach the same detection limits compared to the rest of the sample.  At the distance of NGC 6503, an absolute magnitude limit of $M_V=-6$ mag corresponds to a visual apparent magnitude of 22.6 mag.  At the distance of NGC 1566, the equivalent visual apparent magnitude at $M_V=-8$ mag is also 22.6 mag.  

The visual inspection is performed in order to minimize the contamination within our final cluster catalogs, necessary to produce robust star cluster catalogs.  Each inspected object from the automatic cluster catalog is assigned one of four classifications:  (1) a symmetric, centrally concentrated cluster, usually displaying a homogeneous color; (2) a concentrated cluster with elongated density profiles and less symmetric light distribution, usually displaying a homogeneous color; (3) a less-compact, multiple peaked system that is blue in color, on top of diffuse light; or (4) a spurious detection such as a foreground/background source, single bright star, bad pixel, or a source that lies too close to the edge of the chip.  Class 4 objects are contaminants and excluded from the final cluster catalog.  Table \ref{tab2} lists the median UV and optical colors for each cluster class as well as the compactness, measured by the median CI value, for each galaxy.  In general, class 3 associations exhibit bluer colors in all bands across all galaxies and are marginally more extended compared to class 1 and 2 clusters.

% from /Users/grasha/Research/Thesis/legus_clusters_v13/colors.py
\begin{deluxetable}{lcccc}
\tabletypesize{\footnotesize}
\tablecaption{Color and Compactness of Cluster Classes\label{tab2}} 
\tablecolumns{5}
\tablewidth{0pt}
\tablehead{
\colhead{Class}				& 
\colhead{m$_{UV}$ -- m$_{U}$}		& 
\colhead{m$_{U}$ -- m$_{B}$}			&  
\colhead{m$_{V}$ -- m$_{R}$}			& 
\colhead{CI}   		
\\
\colhead{}		& 
\colhead{(mag)}		& 
\colhead{(mag)}		& 
\colhead{(mag)}		& 
\colhead{(mag)}		
\\
\colhead{(1)}		& 
\colhead{(2)}		& 
\colhead{(3)}		& 
\colhead{(4)}		& 
\colhead{(5)}		
}
\startdata 
\multicolumn{5}{c}{NGC 7793}\\
\hline\\
Class 1 		&	$-$0.29	& $-$0.64	& 0.53	& 1.57(0.18)	\\
Class 2 		&	$-$0.35	& $-$1.12	& 0.57	& 1.57(0.19)	\\  
Class 3 		&	$-$0.46	& $-$1.29	& 0.28	& 1.62(0.21)	\\
\hline\\
\multicolumn{5}{c}{NGC 3738}\\
\hline\\
Class 1 		&	$-$0.14	& $-$0.22	& 0.57	& 1.72(0.21)	\\
Class 2 		&	$-$0.52	& $-$0.62	& 0.47	& 1.67(0.18)	\\  
Class 3 		&	$-$0.57	& $-$0.86	& 0.56	& 1.76(0.11)	\\
\hline\\
\multicolumn{5}{c}{NGC 6503}\\
\hline\\
Class 1 		&	$-$0.36	& $-$0.48	& 0.67	& 1.50(0.17)	\\
Class 2 		&	$-$0.53	& $-$0.67	& 0.51	& 1.53(0.17)	\\  
Class 3 		&	$-$0.56	& $-$0.99	& 0.57	& 1.56(0.23)	\\
\hline\\
\multicolumn{5}{c}{NGC 3344}\\
\hline\\
Class 1 		&	$-$0.21	& $-$0.58	& 0.68	& 1.50(0.12)	\\
Class 2 		&	$-$0.38	& $-$1.28	& 0.40	& 1.51(0.13)	\\  
Class 3 		&	$-$0.50	& $-$1.46	& 0.13	& 1.60(0.16)	\\
\hline\\
\multicolumn{5}{c}{NGC 628}\\
\hline\\
Class 1 		&	$-$0.039	& $-$0.38	& 0.71	& 1.57(0.14)	\\
Class 2 		&	$-$0.20	& $-$0.85	& 0.59	& 1.57(0.14)	\\  
Class 3 		&	$-$0.30	& $-$1.18	& 0.52	& 1.62(0.18)	\\
\hline\\
\multicolumn{5}{c}{NGC 1566}\\
\hline\\
Class 1 		&	0.065		& $-$0.38	& 0.64	& 1.48(0.11)	\\
Class 2 		&	$-$0.21	& $-$1.02	& 0.53	& 1.50(0.14)	\\  
Class 3 		&	$-$0.29	& $-$1.28	& 0.42	& 1.57(0.17)	
\enddata
\tablecomments{
Columns list the 
(1) Classification of stellar clusters; 
(2) median UV color $m_{\rm F275W} - m_{\rm F336W}$; 
(3) median optical color $m_{\rm F336W} - m_{\rm F435W}$; 
(4) median optical color $m_{\rm F555W} - m_{\rm F814W}$; and 
(5) median concentration index.  
Numbers in parentheses indicate uncertainties in the final digit(s) of listed quantities, when available. 
}
\end{deluxetable}

\begin{figure}
%\epsscale{1.2}
%\plotone{histogram_classes_1.eps}
\includegraphics[scale=.4,bb=0 0 576 432]{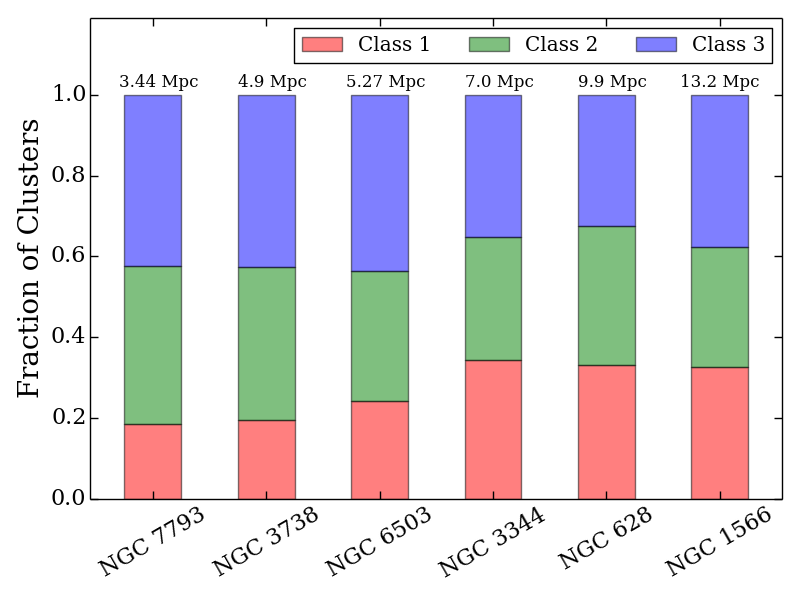}
%python_hist.py
\caption{
Fractional distribution of cluster type within each galaxy, where the color represents the cluster classification and the galaxies are ordered in increasing distance, listed on the top of each bin.  
\label{fig:classhist}}
\end{figure}

The visual classification of each object is performed by at least three independent members of the LEGUS team and the final {\it visually classified} cluster catalog is compiled by comparing all the results from each individual.  Figure \ref{fig:classhist} shows the fractional distribution of cluster types (classes 1, 2, and 3) within each galaxy.  The observed increase in the fraction of class 1 clusters, and, to a minor extent, of both class 1 and 2 relative to the total, for increasing distance may be an effect of decreasing spatial resolution at larger galaxy distance as clusters are not fully resolved and our ability to recover their morphology is dependent on both the distance of the galaxy and the severity of crowding where the individual clusters are located.  However, this is a small effect, especially when class 1 and 2 sources are considered together and compared with the frequency of class 3 sources.

The main difference between the classifications is linked to the morphology of the systems.  Classifications 1 and 2 are considered to be compact star clusters candidates, and for brevity, we will refer to these as star clusters throughout this paper \citep{adamo17}.  It is very challenging to constrain the dynamical state of young stellar clusters; we are not able to determine whether or not the star clusters are actually gravitationally bound.  Class 3 objects may constitute a different type of source compared to class 1 and 2 clusters, where their multi-peak and asymmetrical morphology suggest that they are likely stellar associations, dissolving on short time scales due to being gravitationally unbound \citep{gielesportegieszwart11}; we refer to class 3 sources as `associations' throughout this work.  Due to the nature in the difference of the physical properties (see Section \ref{sec:agehist} and \ref{sec:masshist}) of class 1 and 2 clusters versus the multi-peak associations, throughout our analysis we quite often examine the (possibly bound) class 1 and 2 clusters together and compare their clustering results separately to that of the class 3 associations.  

Detailed descriptions of the completeness tests, applied to NGC 628 as a test source, and how the impact of the assumptions and the selection criteria affect the LEGUS cluster catalogs can be found in \citet{adamo17}.  In brief, at the distance of NGC 628 (9.9 Mpc) and a CI cutoff of 1.4 mag for the central pointing, we are unable to resolve very compact clusters with a size of $R_{\rm eff}=1$ pc and smaller.  We are not worried about missing clusters in our analysis as the size distribution of clusters within NGC 628 peaks at values around 3 pc \citep{ryon17}, well above our completeness limit of 1 pc.  The magnitude limits corresponding to 90\% recovery range from 22.6 to 24.3 mag, depending on the band.  Additionally, clusters older than $\sim$300 Myr are affected by incompleteness due to evolutionary fading, however, only the very youngest clusters $\lesssim 100$ Myr have the strongest influence on the clustering results, and thus, we are not heavily concerned with incompleteness affecting the results and analysis.  We investigate selection effects due to stochastic sampling of the IMF and the impact on the derived cluster ages in Section \ref{sec:selection} and the impact on the clustering results and analysis when applying mass cuts to the sample.

\subsection{Age Distribution}\label{sec:agehist}
\begin{figure}
%\epsscale{1.2}
%\plotone{conf_hist.eps}
%\plotone{conf_hist1.eps}
\includegraphics[scale=0.42]{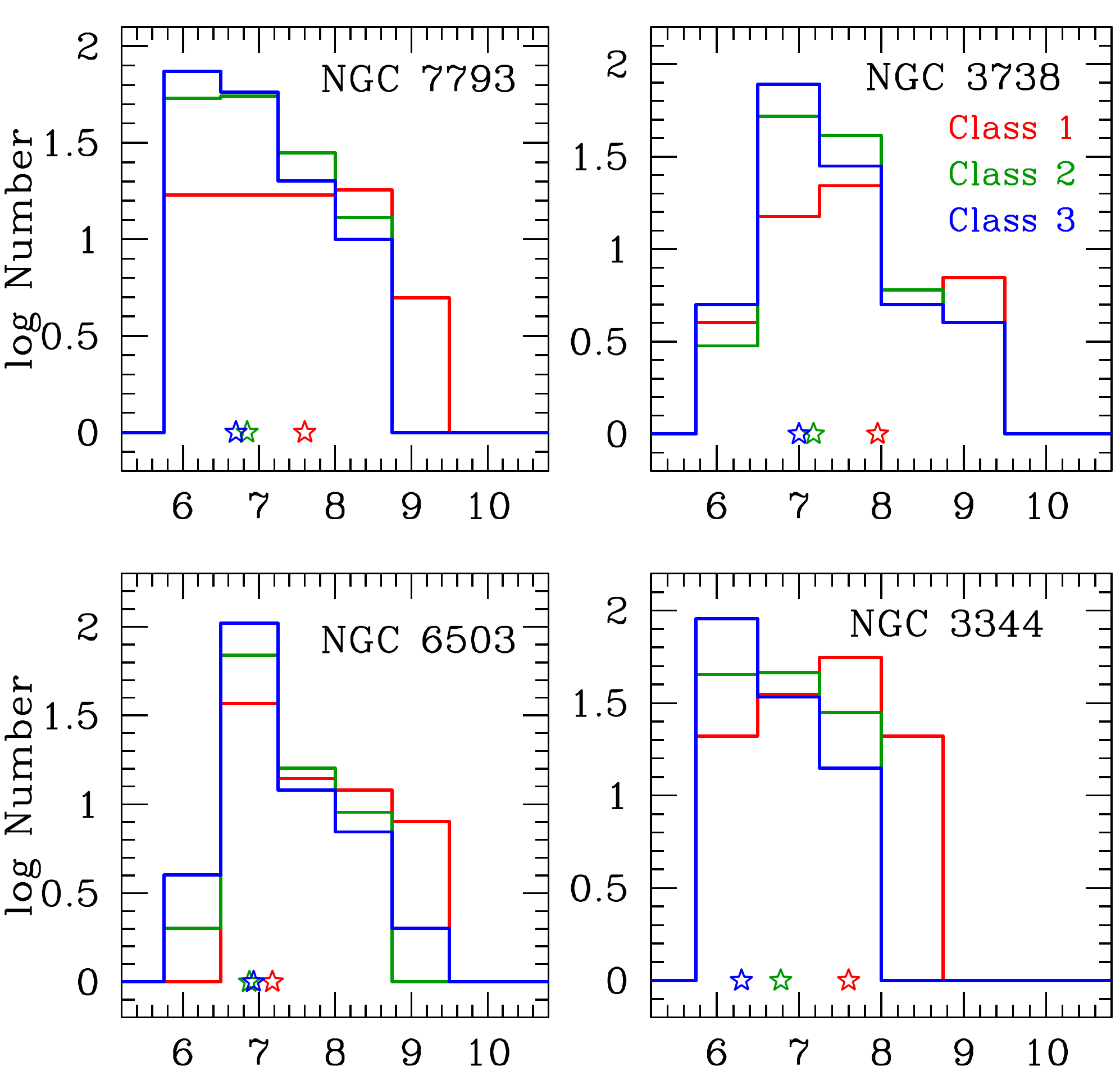}\\
\includegraphics[scale=0.42]{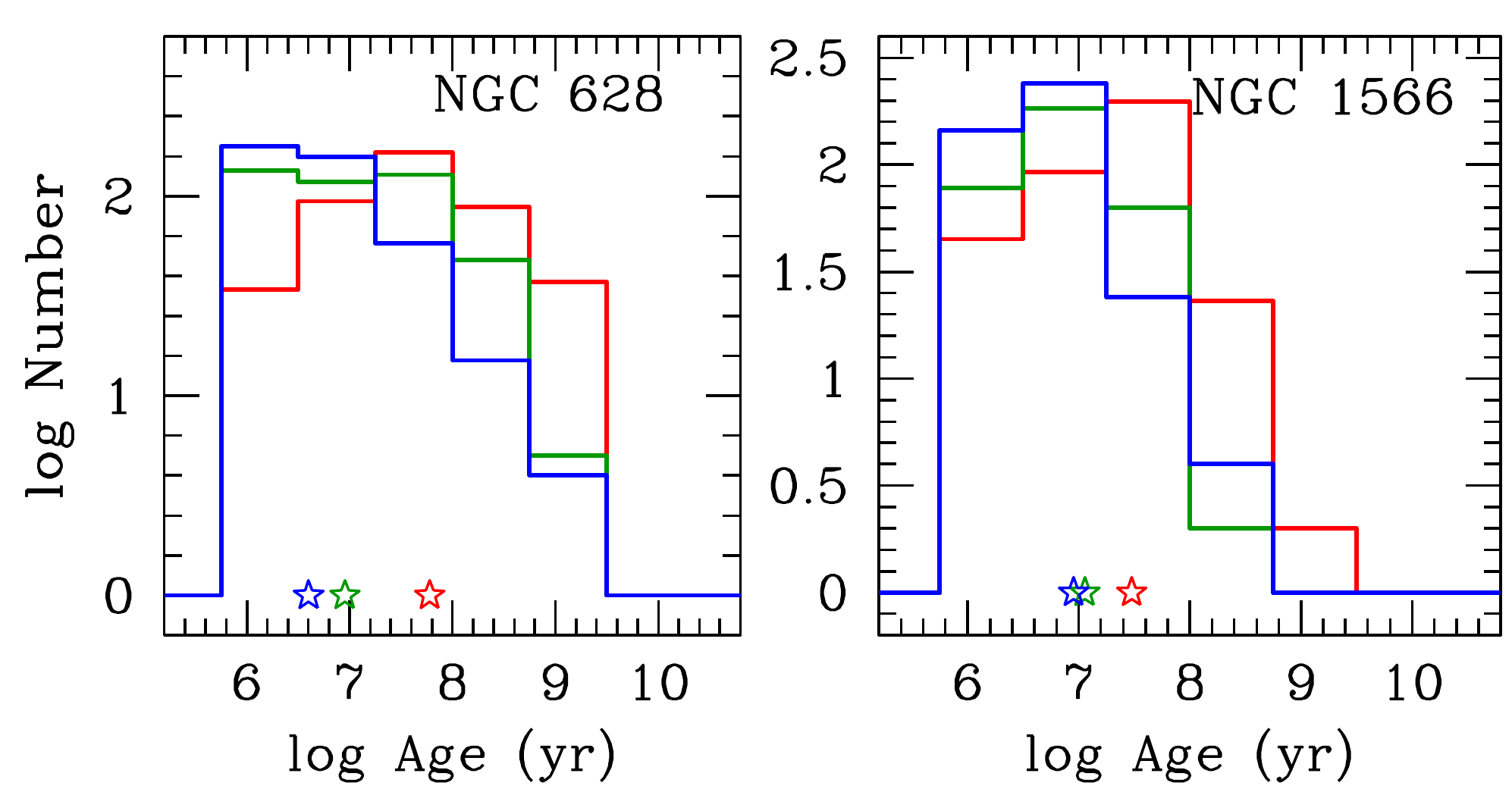}
\caption{
Distribution of ages for each cluster class in each of the galaxies, separated by class type: class 1 (red; symmetrical), 2 (green; asymmetrical), and 3 (blue; multiple peak).  The open star symbol shows the median age of each distribution.  Note the apparent deficit of truly young sources in both NGC 3738 and NGC 6503.
\label{fig:hist}}
\end{figure}

Figure \ref{fig:hist} shows the distribution of the stellar cluster ages for each galaxy broken down by cluster classification along with the median age.  Class 1 clusters (symmetrical) have the oldest median age and the median age of the population decreases for class 2 (asymmetrical) clusters and class 3 (multiple peak) associations, respectively.  A notable feature is the flat or decreasing distribution toward larger ages, especially visible in the class 3 associations, consistent with cluster disruption on timescales of a few tens of Myr or less; cluster fading can compound this effect as well.  The class 3 population does behave distinctly differently compared to class 1 and 2 clusters: in addition to a decrease in their numbers at ages older than a few 10 Myr, their clustering properties in NGC 628 behave quite differently than class 1 and 2 clusters \citep{grasha15}, and their disruption rate is significantly higher compared to class 1 and 2 clusters \citep{adamo17}.  Thus, we maintain that the class 3 associations represent a distinct type of system and appear to be born with lower densities than class 1 and 2 clusters and are not the remnants of dispersed/disrupted class 1 or 2 clusters.  %Overall, the age distributions of the clusters are roughly flat for ages longer than a few Myr, consistent with most clusters disrupting on timescales of a few tens of Myr or less.  

\subsection{Mass Distribution}\label{sec:masshist}
\begin{figure}
\epsscale{1.2}
\includegraphics[scale=0.42]{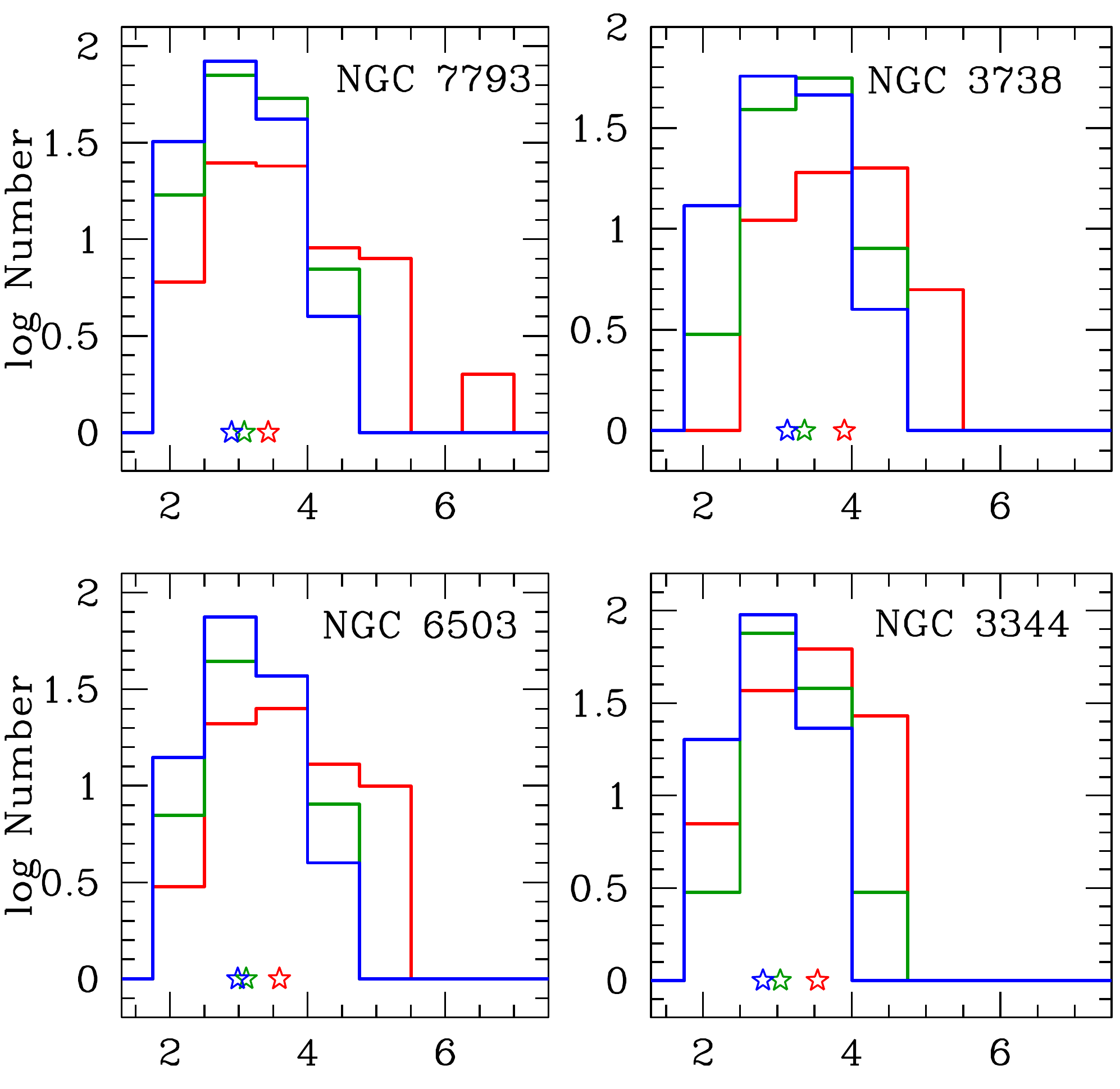}\\
\includegraphics[scale=0.42]{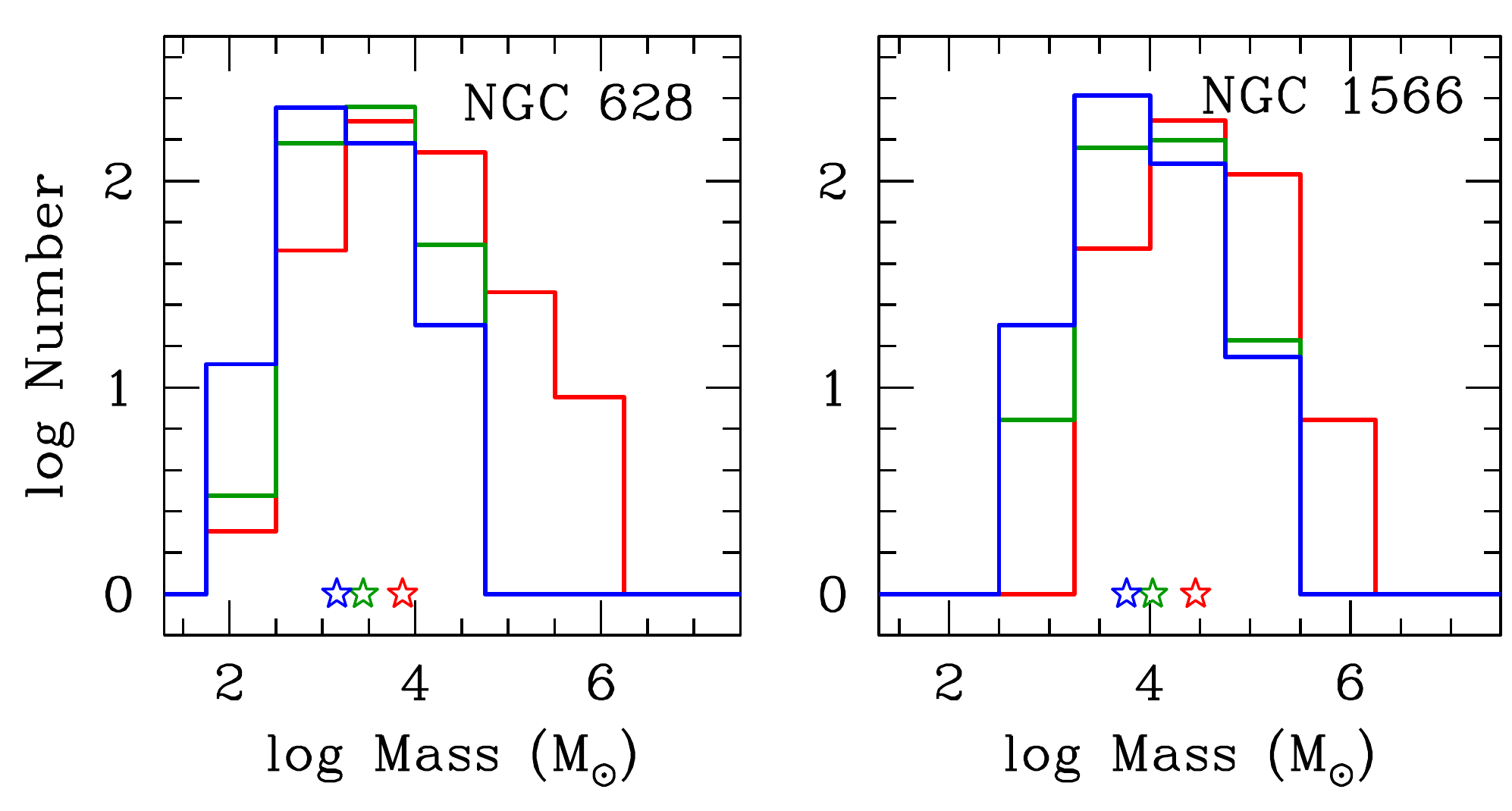}
\caption{
Distribution of mass for each class galaxy, separated by class type: Class 1 (red; symmetrical), 2 (green; asymmetrical), and 3 (blue; multiple peak).  The open star symbol shows the median mass of each distribution.  
\label{fig:histmass}}
\end{figure}

Figure \ref{fig:histmass} shows the distribution of the cluster masses within each galaxy, divided by cluster classification.  The clusters display mass trends across all six galaxies that are dependent on their classification, similar to what is observed for the age distributions (Section \ref{sec:agehist}): class 1 clusters, on average, are more massive compared to the rest of the clusters and the class 3 associations are the least massive.  The three morphological categories of cluster candidates, once we take into account their different age and mass properties, also reinforces the physical differences between the classifications, as discussed in Section \ref{sec:agehist}.  Class 1 clusters are older and more massive stellar clusters, indicative of evolved systems that are relaxed and have likely already survived disruption.  Class 3 associations, on the other hand, are clusters that are much younger and less massive compared to bound stellar clusters.  

\subsection{Deprojection of the Galactic Disk}\label{sec:deproj}
In order to assess the spatial distribution of the young stellar clusters free from the effect of projection of the galactic disk.  We deproject the stellar cluster positions from the plane of the sky to the plane of the galaxy using the simple assumption that each galaxy can be described with an axisymmetric flat rotating disc.  

\begin{figure*}      
\includegraphics[scale=.45, bb=0 0 500 500]{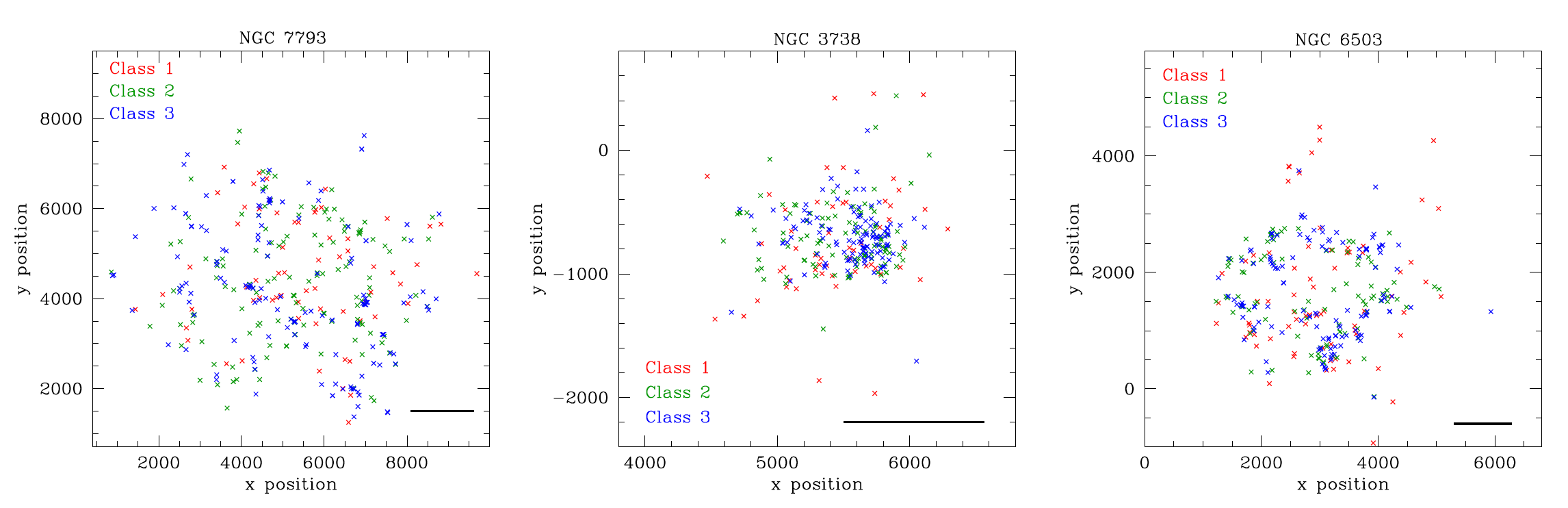}
    \caption{The deprojected pixel positions of the stellar clusters after taking into account the inclination angle, color coded by the classification of the clusters:  class 1 clusters are shown in red, class 2 clusters are shown in green, and class 3 associations are shown in blue.  NGC 628, NGC 3344, and NGC 1566 are not corrected for inclination as the deprojection is a small effect for these galaxies.  The solid black line in the bottom right of each figure represents the spatial scale of 1 kpc at the distance of that galaxy. }
    \label{fig:nodeproj}
\end{figure*}

The positions of star clusters are corrected for the inclination $i$ of the galaxy in a two step process.  First the intrinsic $x,y$ detector coordinates in the plane of the galaxy are rotated as
\begin{equation}
\begin{gathered}
x' = x \cos \theta + y \sin\theta \\ 
y' = -x \sin \theta + y \cos \theta,
\end{gathered}
\end{equation}
where $x',y'$ are the rotated coordinate axes and the position angle $\theta$ is determined by the orientation of the observed field of view.  The rotated pixel coordinates are then deprojected for the line of sight inclination angle as $x'$ and $y'/\cos i$.  We have found that for galaxies with an inclination angle below $40^{\circ}$, correction for projection has a minimal impact on the relative position of the clusters, and therefore, does not influence the clustering results.  Figure \ref{fig:nodeproj} shows the spatial distributions of the stellar clusters, where the positions of the clusters in three galaxies with $i>40^{\circ}$ have been deprojected, NGC 7793, NGC 3738, and NGC 6503.  For all calculations in this paper, we use the deprojected positions of the star clusters for the above three galaxies.  Due to the low inclination of NGC 628, NGC 3344, and NGC 1566, the star cluster positions are not corrected for projection effects and we use the cluster positions as shown in Figure \ref{fig:gal}.  

\section{The Two-Point Correlation Function}\label{sec:2pcf}
We implement the angular two-point correlation function, $\omega(\theta)$, as projected onto the plane of the sky, to measure the magnitude of clustering as a function of scale size for the young stellar clusters in six nearby galaxies (Table \ref{tab1}).  The two-dimensional correlation function $1+\omega(\theta)$ is defined as the probability above Poisson of finding two star clusters with an angular separation $\theta$ as $\mathrm{d}P = N^2 [ 1 +  \omega(\theta)]\ \mathrm{d}\Omega_1 \mathrm{d}\Omega_2$, where $N$ is the surface density of clusters per steradian with two infinitesimal elements of solid angle $\mathrm{d}\Omega_1$ and $\mathrm{d}\Omega_2$, separated by angle $\theta$ \citep{peebles80}.  For truly random Poisson distribution, the two-point correlation function will be flat across all scales, such that $1+\omega(\theta) = 1$ and a clustered stellar sample will have $1+\omega(\theta) > 1$ at small values of $\theta$, declining with increasing scales toward that of a flat, non-clustered distribution.  The correlation function of a fractal (self-similar) distribution is described with a single power-law as $1+\omega(\theta) = (r/r_0)^{-\gamma}$, where $r_0$ is the characteristic scale-length of the clustering and $\gamma$ describes the hierarchical ordering (see Section \ref{sec:powerlawfit}). 

To measure $\omega(\theta)$, pairs of stellar clusters are counted as a function of their separation (deprojected if necessary), compared to what is expected for an unclustered distribution.  The unclustered distribution (in x, y position) of sources must populate the same sky coverage and geometry (e.g., edges, masks) as the observational data.  We define masks as areas that exclude all data, such as the ACS chip gap or areas with dust lanes, where there is a reduction in the observed number of clusters with respect to the global average.  We reproduce, as closely as possible, the geometry of the galaxy region sampled for each random catalog.  The ratio of pairs of clusters observed in the data relative to pairs of points in the random catalog is then used to estimate the correlation function $\omega(\theta)$.  In this study, we implement the Landy-Szalay estimator \citep[LS;][]{landy93}, calculated as, %designed to minimize systematic and random errors
\begin{equation}\label{eq:LS}
\omega_{\rm LS}(\theta) = \frac{DD(\theta) - 2DR(\theta) + RR(\theta)}{RR(\theta)},
\end{equation}
where $DD$ is the number of data-data pairs, $DR$ is the number of cross-correlated data-random pairs, and $RR$ is the number of random-random pairs with the same mean density and sampling geometry with separation between $\theta$ and $\theta + \delta \theta$.  The pair counts are computed as,
\begin{equation}\label{eq:terms}
\begin{gathered}
DD(\theta) = \frac{P_{DD}(\theta)}{N(N-1)/2}\\[6pt]
DR(\theta) = \frac{P_{DR}(\theta)}{N N_{R}}\\[6pt]
RR(\theta) = \frac{P_{RR}(\theta)}{N_{R}(N_{R}-1)/2}, \\[6pt]
\end{gathered}
\end{equation}
where $N$ and $N_R$ are the total number of data and random points in the survey volume, respectively and $P_{DD}(\theta)$, $P_{DR}(\theta)$, and $P_{RR}(\theta)$ represent the total cluster counts in each separation $\theta \pm \delta \theta$ bin for the data-data, data-random, and random-random pairs, respectively.  The size of the bin is determined by the sample size, selected as a compromise between resolution and total number of clusters available to be sampled within each bin.  The formulation in Eq \ref{eq:LS} is optimized to mitigate edge effects and systematic errors on the computed correlation function, necessary to alleviate unrealistic calculations of $\omega(\theta)$ at large spatial scales due to the deviation from a random distribution which can arise from a limited field of view and edge effects.  In \citet{grasha15}, we investigate the influence of the random catalog size and geometry on the outcome of the two-point correlation function.  

\begin{figure*}
%\epsscale{1.1}
%\plottwo{ngc7793_2pcf_mg1.eps}{ngc3738_deproj_2pcf_mg1.eps}
%\plottwo{ngc6503_deproj_2pcf_mg1.eps}{ngc3344_2pcf_mg1.eps}
%\plottwo{ngc628_2pcf_mg1.eps}{ngc1566_all_2pcf_mg1.eps}
\includegraphics[scale=.8, bb= 0 0 100 600]{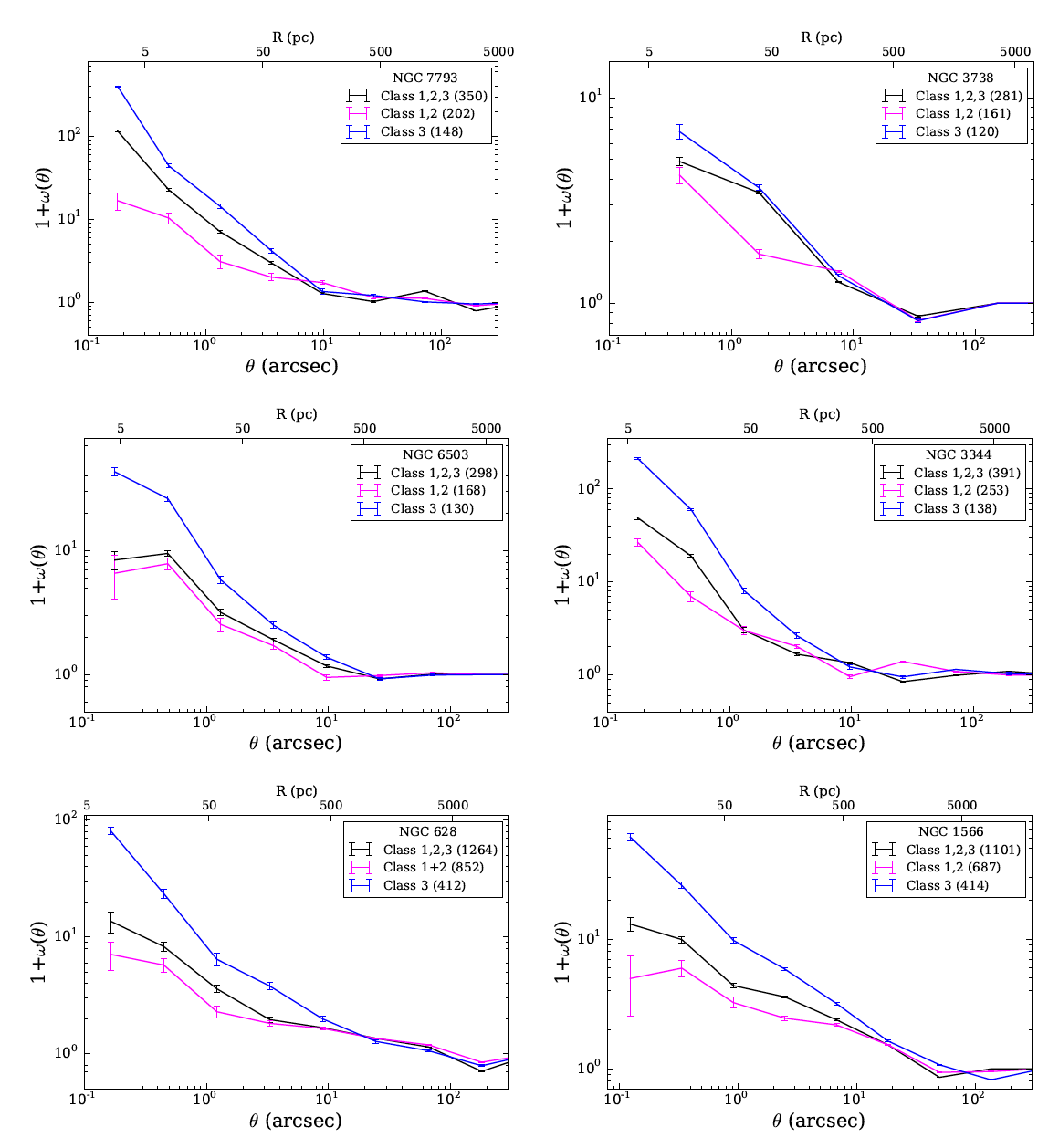}
\caption{
The two-point correlation function $1+\omega(\theta)$ for the clusters as a function of angular distance (arcsec), with the top axis showing the corresponding spatial scale at the distance of the galaxy.  The colors represent the classification of each cluster, as defined in Section \ref{sec:clusterselection}: class 1+2 clusters are shown in magenta, class 3 associations are shown in blue, and all clusters are shown in black.  The numbers in parentheses show the number of clusters in each classification.  The 1$\sigma$ uncertainties are the sample standard deviation of bootstrap resamples within each bin. 
\label{fig:2pcf}}
\end{figure*}

Figure \ref{fig:2pcf} shows the correlation functions for all six galaxies, calculated for the entire sample of star clusters and divided by their cluster classification.  As expected for star-forming structures formed hierarchically, there is a systematic decrease in the clustering with increasing spatial scales across all galaxies.  Additionally, the clustering strength of the class 1+2 (bound) clusters across all the systems is much weaker than what is observed for the class 3 associations alone.  Thus, the bound clusters -- for all ages -- have a more homogeneous and relaxed distribution throughout their galaxy at a given spatial scale compared to the (highly clustered) distribution of associations.  This can result from either different formation/evolutionary paths between bound and unbound clusters and/or an age dependency of the clustering strength, as found in \citet{grasha15} and investigated further in Section \ref{sec:agecuts} and Section \ref{sec:combined} for these systems.  
%We will discuss in Section \ref{sec:powerlawfit} and \ref{sec:agecuts} the differences between the correlations for the separate classes of clusters.  
%These results indicate that bound clusters (class 1 and 2) are less clustered than class 3 and of the combination of class 1, 2, and 3 clusters together.  
%This reinforces our earlier result from NGC 628 alone \citep{grasha15} that as (bound) clusters relax, they become less clustered and behave differently than what is observed for associations.  

\subsection{Selection Effects}\label{sec:selection}
At the distance of NGC 628 (9.9 Mpc), the LEGUS cluster catalogs are complete down to 5000 M$_{\odot}$ \citep{adamo17}.  As outlined in Section \ref{sec:clusterselection}, we do not consider stochastic sampling of the IMF when deriving the mass and age of the clusters, which becomes increasingly important for clusters with masses below $\sim 5000$ M$_{\odot}$.  However, as found in simulations by \citet{krumholz15}, the properties of the cluster population as a whole are relatively similar between both conventional deterministic and stochastic fitting procedures.  A comparison of the cluster ages show that there is no difference in the results for ages that are derived with stochastic versus non-stochastic models \citep{grasha17}.  As the clustering results is most heavily influenced by the youngest clusters \citep[$<$100 Myr; see Section \ref{sec:agecuts} as well as][]{grasha15}, which may lie below the mass limit of $5000$ M$_{\odot}$ \citep{adamo17}, it is important to examine how mass-cuts influence the clustering analysis to ensure that our results are not biased against the detection limitations. %of low-mass clusters.  

\begin{figure}
\epsscale{1.2}
\plotone{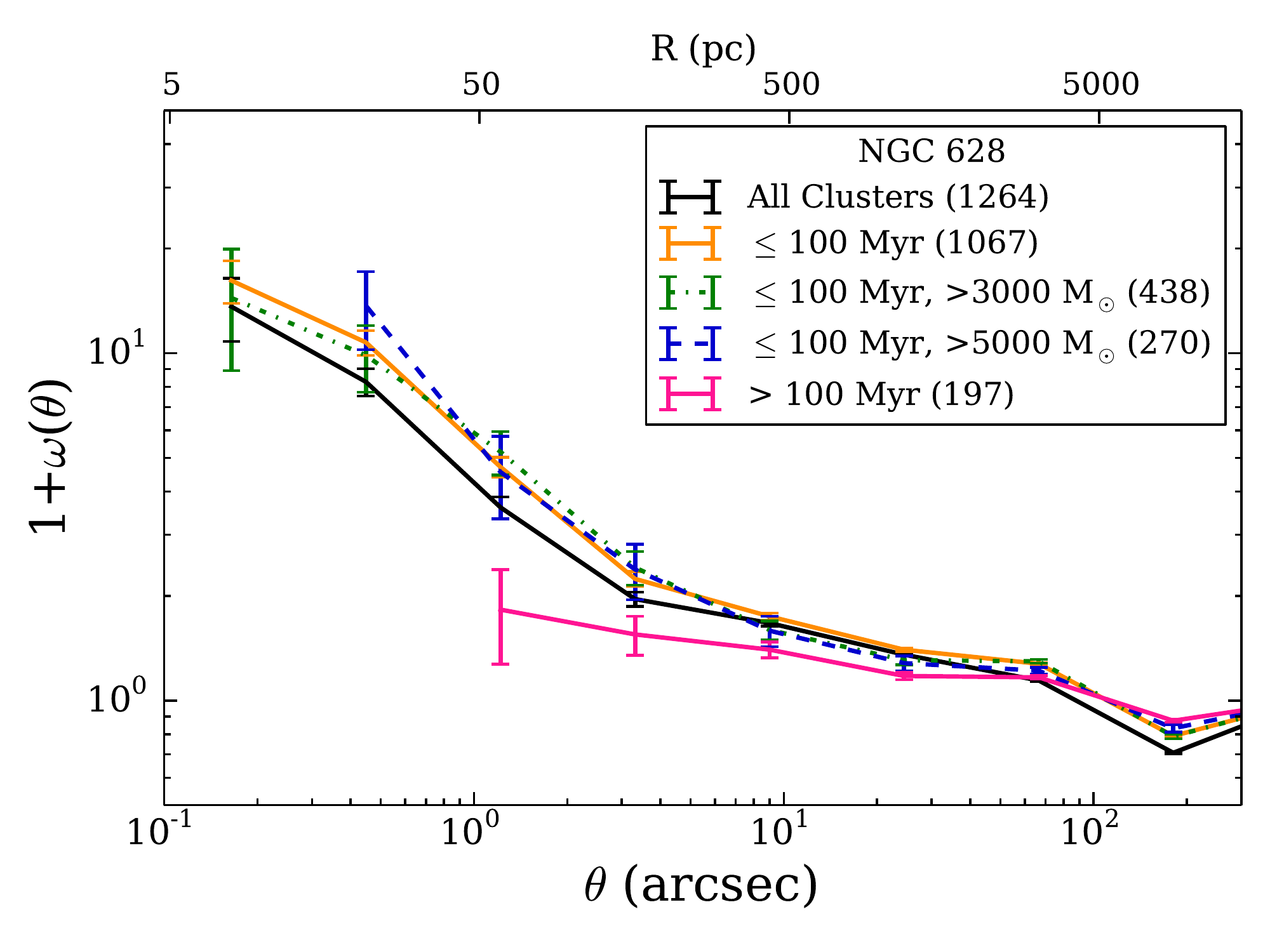}
%BoundingBox:  18 144 575 702
% from plottinginpython_ngc628_masscut.py
\caption{
The two-point correlation function $1+\omega(\theta)$ of the clusters in NGC 628 as a function of angular distance (arcsec), with the top axis showing the corresponding spatial scale, showing the effect that mass cutoff has on the clustering results.  The colors represent different age and mass ranges and the numbers in parentheses show the number of clusters in each age/mass bin.  For all clusters with ages less than 100 Myr, the correlation function does not change for different mass cuts.  Thus, we are not worried about selection effects affecting the clustering results and analysis.  However, there is a pronounced decrease in the clustering strength for clusters older than 100 Myr; this effect of age on the clustering strength is discussed in Section \ref{sec:agecuts}.  
\label{fig:masscut}}
\end{figure}

Figure \ref{fig:masscut} shows the clustering correlation for the young ($<100$ Myr) clusters in NGC 628 with mass cuts applied at both 3000 and 5000 M$_{\odot}$.  For both cases, the clustering strength does not change when applying a mass cut and the correlation functions are in agreement with each other and with the entire cluster population.  The reason there is no change to the observed correlation function for high/low mass clusters is easily understood if we consider that low-mass clusters ought to be populated throughout the galaxy in a similar manner as the entire cluster population.  Consequently, incompleteness effects will not manifest themselves in a way that missing clusters (in this case, low-mass clusters) are preferentially located within one region of a galaxy, affecting the global clustering distribution.  Missing low-mass clusters evenly throughout a galaxy only acts to reduce the total clusters available for analysis, and if anything, serves to artificially reduce the observed global clustering.  Due to this negligible effect, we do not perform a mass cut on any galaxy and have no need to worry about selection effects affecting the clustering results.  

\section{Results and Analysis}\label{sec:results}

\subsection{Quantifying the Correlation Strength}\label{sec:powerlawfit}
In a two-dimension (projected) self-similar distribution, the total number of clusters $N$ increases with radius $r$ as $N \propto r^{D_2}$, where $D_2$ is the two-dimensional fractal dimension \citep{mandelbrot82}.  Fractal distributions exhibit a power-law dependency of the correlation function $1+\omega(\theta)$ with increasing radius of the form $1+\omega(\theta) \propto r^{\alpha}$ \citep{calzetti89,larson95} and the number of clusters for every radial aperture will increase as $N \propto r^{\alpha} \times r^{2} \propto r^{\alpha +2}$.  Thus, we can then relate the power-law slope $\alpha$ from the correlation function $1+\omega(\theta)$ to determine the two-dimensional fractal dimension as $D_2 = \alpha + 2$.  A flat, non-clustered distribution of $\alpha=0$ will result in a fractal geometric dimension of $D_2=2$; a derived steep (negative) slope will be indicative of a clustered distribution, resulting in a fractal dimension less than 2.  We determine the slope $\alpha$ from the correlation function $1+\omega(\theta)$ at small scales in log-log space through a Levenberg-Marquardt non-linear least square minimization fit as 
\begin{equation}\label{eq:powerlaw}
1+\omega(\theta) = A_{\omega}\theta^{\alpha},
\end{equation}
where the slope $\alpha$ measures the strength of the clustering and $A$ measures the amplitude of the clustering.  Often, the correlation function is best described with a double power law as opposed to a single slope \citep[e.g., ][]{larson95,gouliermis14}, determined through minimizing $\chi^2$.  For those cases, we determine two slopes, using the functional form is given as, 
\begin{equation}
\log [1+\omega(\theta)] = \left\{
  \begin{array}{lr}
    A_1 + \alpha_1 \log(\theta) &  : \log(\theta) < \beta \\
    A_2 + (\alpha_1 - \alpha_2)\beta  + \alpha_2 \log(\theta) &  : \log(\theta) > \beta,
  \end{array}
\right.
\end{equation}
where the breakpoint $\beta$ is the logarithm of the position of the separation break along the x-axis (spatial scale), $A_1$ and $A_2$ are the clustering amplitudes before and after the break, and $\alpha_1$ and $\alpha_2$ are the slopes of the power law before and after the breakpoint, respectively.  Both slopes and the breakpoint are free parameters in the fit.  %The fitted parameter results are listed in Table \ref{tab3}.  

For a hierarchical model, the distribution of star formation and interstellar gas, over a large range of environments, is shown to exhibit a projected, two-dimensional fractal dimension of $D_2\sim$ 1.2--1.6 \citep[e.g., ][]{beech87,falgarone91,elmegreen96f,sanchez05,elmegreen06,sanchez08}.  Smaller values of the fractal number $D_2$ correspond to higher fractal dimensions (i.e., systematically more clustering) and steeper slopes for the correlation function.  These values for the 2D fractal dimension are in agreement with the predicted fractal dimension for the three-dimensional, $D_3$, density structure of the ISM \citep{federrath09} in a turbulent medium.  Often, the connection between the three-dimensional fractal dimension $D_3$ and the projected two-dimensional fractal $D_2$ is assumed to be $D_2$ = $D_3$ -- 1 \citep{mandelbrot82}.  However, the conversion between $D_2$ and $D_3$ is not straightforward \citep{gouliermis14a}.  %, and in general, the simple conversion is not necessarily valid.  

Table \ref{tab3} lists the distribution of power law slopes and amplitudes for all the clusters in each galaxy.  %Class 1+2 clusters cover a narrow range of slopes ($\sim$ --0.1 to --0.9) and an even narrower range in amplitudes ($\sim$ 2--6) compared to recovered slopes of class 3 associations ($\sim$ --0.4 to --1.6) and amplitudes ($\sim$ 3--27).  Class 3 associations also exhibit an increase in their amplitude when the recovered slope is steeper.  
All galaxies are best-fit with a single power law, excluding NGC 628 and NGC 3344, which are best-represented by a double power law.  We fit over the dynamical range up to the scale where the correlation function becomes flat ($R_0$, the correlation length) as values of $1+\omega(\theta)=1$ corresponds to a random (non-clustered) distribution.  Similar breaks in the power-law have been observed in both the correlation function of young stars in M31 \citep{gouliermis14} and in the spectral correlation function of the Large Magellanic Cloud \citep[LMC;][]{padoan01}, attributed to the disk scale height and corresponding to the transition between turbulent motions at small scales to disk dynamics at large scales.  
%For three of our galaxies (NGC 628, NGC 3344, NGC 7793), the correlation function for certain subsamples of clusters (e.g. Class 3, Class 1+2, etc.) is best described with a double power law.

%, i.e, NGC 1566 is fit from 0.13'' (11 pc) to 50" (4400 pc) whereas NGC 6503 is only fit from 0.18'' (5 pc) to 26'' (733 pc).  

\begin{deluxetable*}{lcccccc}
\tabletypesize{\scriptsize}
\tablecaption{Power-Law Parameters for Individual Galaxies \label{tab3}} 
\tablecolumns{7}
\tablewidth{0pt}
\tablehead{
\colhead{Class}				& 
\colhead{Number}			& 
\colhead{$A_1$}				&  
\colhead{$\alpha_{1}$} 	&
\colhead{$\beta$}   		&
\colhead{$A_2$}				&  
\colhead{$\alpha_2$} 	
\\
\colhead{}		& 
\colhead{}		& 
\colhead{}		&  
\colhead{} 		&
\colhead{('')}   &
\colhead{} 		&
\colhead{}   
}
\startdata 
\multicolumn{7}{c}{NGC 7793}\\
\hline\\
%Class 1,2 	&	202		& 3.8(3)	&  $-0.91(7)$ &  1.3		& 3.2(2) & $-$0.25(2) \\
%Class 3 		&	148		& 26.8(8)	&  $-1.54(2)$ &  		&	&  \\
Class 1,2,3 		&	350		& 14.0(6)	& 	$-1.16(5)$ & \multicolumn{3}{c}{\nodata}\\ 
\hline\\
\multicolumn{7}{c}{NGC 3738}\\
\hline\\
%Class 1,2 	&	161		& 2.75(6)	&  $-0.34(3)$ &  		&	&\\
%Class 3 		&	120		& 3.86(9)	&  $-0.46(3)$ &  		&	& \\
Class 1,2,3 		&	281		& 3.27(3)	& 	$-0.394(12)$ & \multicolumn{3}{c}{\nodata}\\
\hline\\
\multicolumn{7}{c}{NGC 6503}\\
\hline\\
%Class 1,2 	&	168		& 3.9(3)	&  $-0.63(6) $ &  		& &  \\
%Class 3 		&	130		& 10.4(2)	&  $-0.94(2)$ &  		& 	&  \\ 
Class 1,2,3 		&	298		& 3.97(19)	& 	$-0.46(6)$ & \multicolumn{3}{c}{\nodata}\\
\hline\\
\multicolumn{7}{c}{NGC 3344}\\
\hline\\
%Class 1,2 	&	253		& 5.4(3)	&  $-0.78(4)$ &  		& 	&  \\
%Class 3 		&	138		& 20.5(6)	&  $-1.356(18)$ &  		& 	& \\
Class 1,2,3 		&	391		& 4.7(3)	& 	$-1.68(13)$ & 	1.3	& 3.37(13) & $-$0.42(3)\\
\hline\\
\multicolumn{7}{c}{NGC 628}\\
\hline\\
%Class 1,2 	&	852		& 2.49(4)	&  $-0.86(7)$ & 1.2		& 2.16(14)	& $-0.14(2)$ \\
%Class 3 		&	412		& 6.11(3)	&  $-1.51(6)$ & 1.2 		& 4.89(3)	& $-0.37(2)$ \\
Class 1,2,3 		&	1264	& 4.81(5)& $-0.75(4)$ 	& 3.3 		& 2.39(5)	& $-0.16(2)$ \\
\hline\\
\multicolumn{7}{c}{NGC 1566}\\
\hline\\
%Class 1,2 	&	431		& 3.76(9)	&  $-0.25(2)$ &  		&	&  \\
%Class 3 		&	234		& 15.3(2)	&  $-0.808(13)$ &  		&	& \\
Class 1,2,3 		&	1101	& 6.06(6)	& 	$-0.492(13)$ & \multicolumn{3}{c}{\nodata}
\enddata
\tablecomments{
Power law fits for separations measured in angle ($\theta$).  Columns list the 
(1) Classification of stellar clusters; 
(2) Number in each classification; 
(3) Amplitude $A_1$ of the angular correlation function before the breakpoint; 
(4) Slope $\alpha_1$ of the angular correlation function before the breakpoint; 
(5) Location of the break point $\beta$.  Cluster classifications that are best-fit with a single power law do not have a breakpoint; 
(6) Amplitude $A_2$ of the angular correlation function past the breakpoint; and
(7) Slope $\alpha_2$ of the angular correlation function past the breakpoint.  Numbers in parentheses indicate uncertainties in the final digit(s) of listed quantities, when available. 
}
\end{deluxetable*}

\subsection{Age Effects}\label{sec:agecuts}
The general distribution of stellar clusters across all the galaxies displays a fairly consistent picture:  class 3 associations are very strongly clustered at the smallest spatial scales (Figure \ref{fig:2pcf}) and in general, have  a younger median age distribution in comparison to the class 1 and 2 clusters (Figure \ref{fig:hist}).  %This differs from what we find for class 1 and 2 clusters, which are possibly bound systems, and exhibit weaker clustering in addition to older median ages.  
In order to investigate an age-dependency to the star-forming structures and the timescale over which the clustered substructures disperse, we divide the clusters by age to recover where the difference in clustering is maximal.  We then recompute the correlation functions in these different age bins, determined individually within each galaxy, to explore how the age influences both the size and the clustering strength of the star-forming complexes within each galaxy.  

% all from plottinginpythong_ngc*.py
\begin{figure*}
%\epsscale{1.1}
\includegraphics[scale=.8, bb= 0 0 100 650]{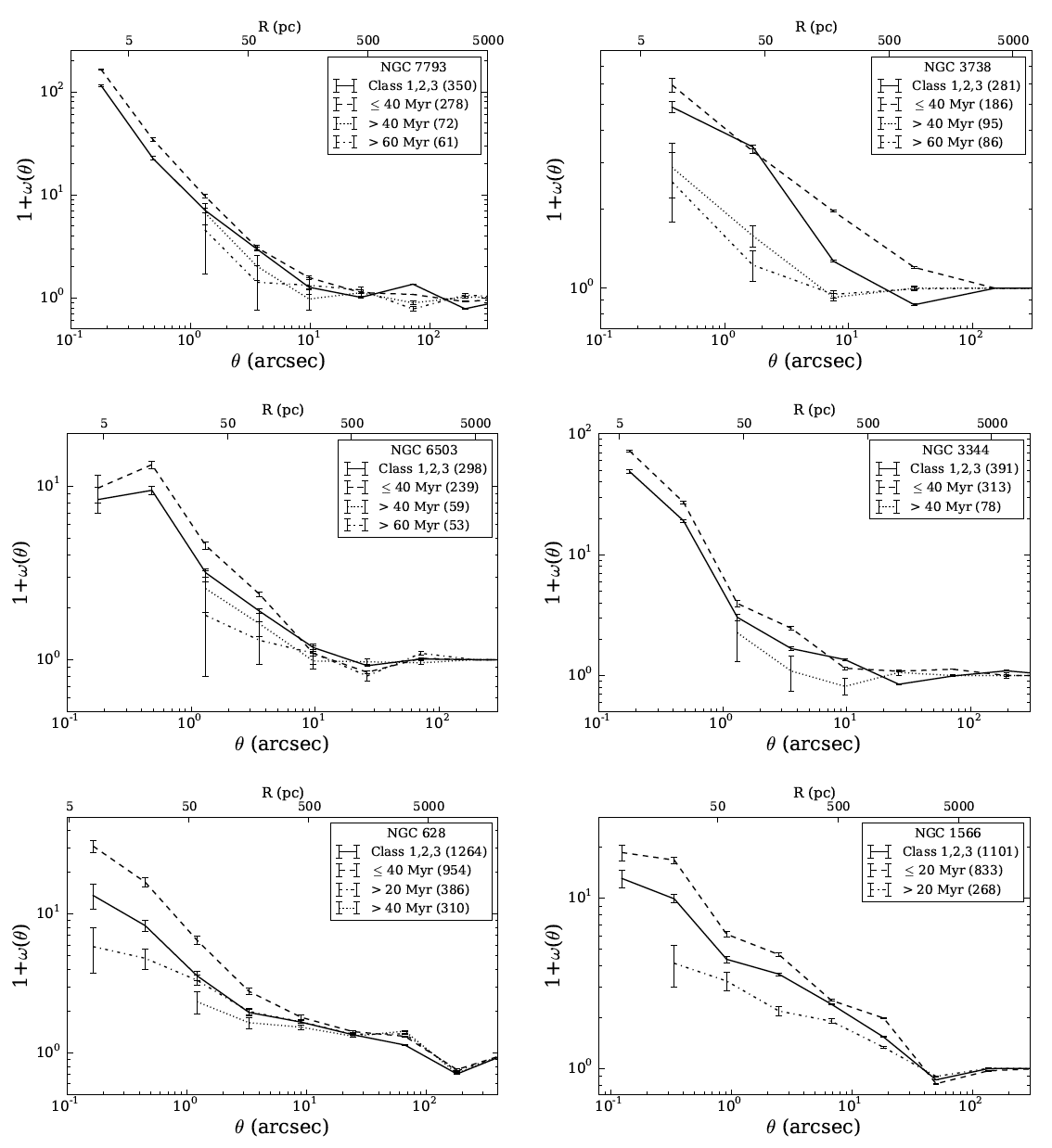}
%\plottwo{ngc7793_deproj_2pcf_agecut_class123.eps}{ngc3738_deproj_2pcf_agecut_class123.eps}
%\plottwo{ngc6503_deproj_2pcf_agecut_class123.eps}{ngc3344_2pcf_agecut_class123.eps}
%\plottwo{ngc628_2pcf_agecut_class123.eps}{ngc1566_2pcf_agecut_all_class123.eps} %ngc1566_2pcf_agecut_class123.eps
%BoundingBox:  18 144 575 702
\caption{
The two-point correlation function $1+\omega(\theta)$ as a function of angular distance (arcsec) for all the cluster classifications as shown in Figure~\ref{fig:2pcf} along with the subset of clusters with ages below/above specific age divisions determined by requiring that the clustering among the young star clusters is maximized.  The strength of clustering increases when we consider the youngest clusters within each classification and affects the smallest spatial scales first within all galaxies.  
\label{fig:2pcfage}}
\end{figure*}

Figure \ref{fig:2pcfage} shows how the clustering strength depends on the age of the clusters.  For all galaxies and cluster classifications, there is a clear dichotomy between younger and older clusters, with clustering decreasing in strength for clusters older than 20--60 Myr.  This timescale for the transition from a clustered distribution of the clusters at the youngest ages to a more smooth, homogeneous distribution occurs very rapidly, and in agreement with similar studies where the observed clustered distribution is slowly lost as the clusters/stars age. \citep[e.g., ][]{scheepmaker09,sanchez09,gouliermis15}.

While small numbers for most of the galaxies is a limiting factor in our analysis, NGC 3738 shows the most striking influence of age on the clustering strength, where there is a dramatic decrease in the observed clustering for all the clusters within the irregular galaxy.  NGC 628 and NGC 1566 also exhibit a substantial decrease in clustering with increasing cluster ages.  NGC 7793, NGC 6503, and NGC 3344 only exhibit a slight change in the clustering with age, and within each of the cases, there are a lack of any older ($\gtrsim$50~Myr) clusters at the smallest spatial scales.  Thus, the amount of clustering occurring at small spatial scales is very uncertain for the older clusters in most of the galaxies.

\subsection{The Effect of Global Galactic Properties on the Correlation Length $R_0$}\label{sec:spatial}
\begin{figure*}
%\epsscale{1.2}
\includegraphics[scale=.5]{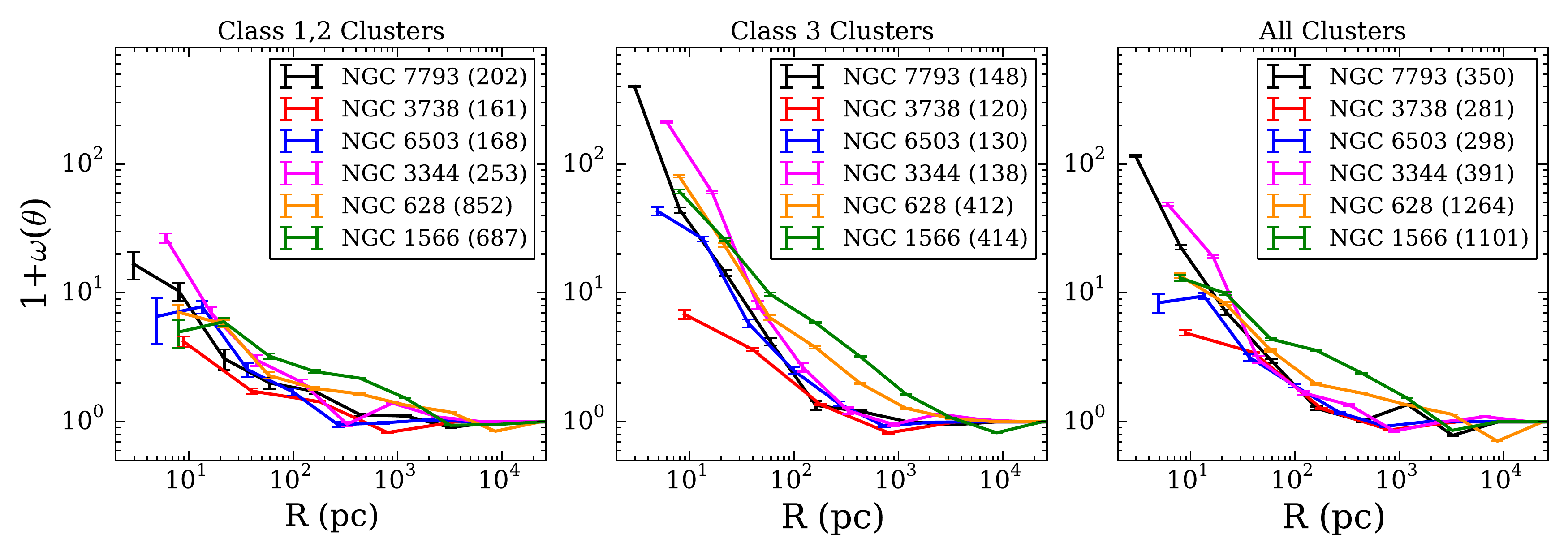}
% made in plottinginpython_master.py
%BoundingBox:  18 144 575 702
\caption{
Combined two-point correlation functions $1+\omega(\theta)$ for every galaxy as a function of physical scale (pc) for the class 1 and 2 clusters (left), the class 3 associations (middle), and all clusters (right).  The numbers in parentheses show the number of clusters in each classification.  
\label{fig:2pcfall}}
\end{figure*}

To compare the clustering results between the galaxies, we convert the length scales in arseconds to a spatial scale (parsec). 
We see in Figure \ref{fig:2pcfall} that the strength of the clustering for a fixed physical scale within each galaxy reflects the difference in structure of the galaxy, where bigger/brighter galaxies are associated with a larger amplitude for a fixed distance between cluster pairs.  The larger clustering amplitude of NGC 1566 compared to the rest of the sample is quite possibly observational and not physical, and much of the discrepancy may be due to adopting a distance that is further than assumed as the distance to the galaxy is quite uncertain.  

To investigate the individual galactic properties on the clustering results, we compare the correlation length $R_0$ to different physical properties of each galaxy for all cluster classifications.  The correlation length $R_0$ is defined where the correlation function flattens ($\omega =0$), i.e. the distributions of star clusters become uncorrelated.  The correlation length $R_0$ can describe the size of the typical star-forming complexes within a galaxy where the youngest star clusters were born and still reside.  Thus, in addition to $R_0$ having an age-dependency, the correlation length may also depend on global galactic properties or any processes that influence the global star formation within these clustered star-forming structures.  

\begin{figure}
%\epsscale{1.2}
\includegraphics[scale=.35]{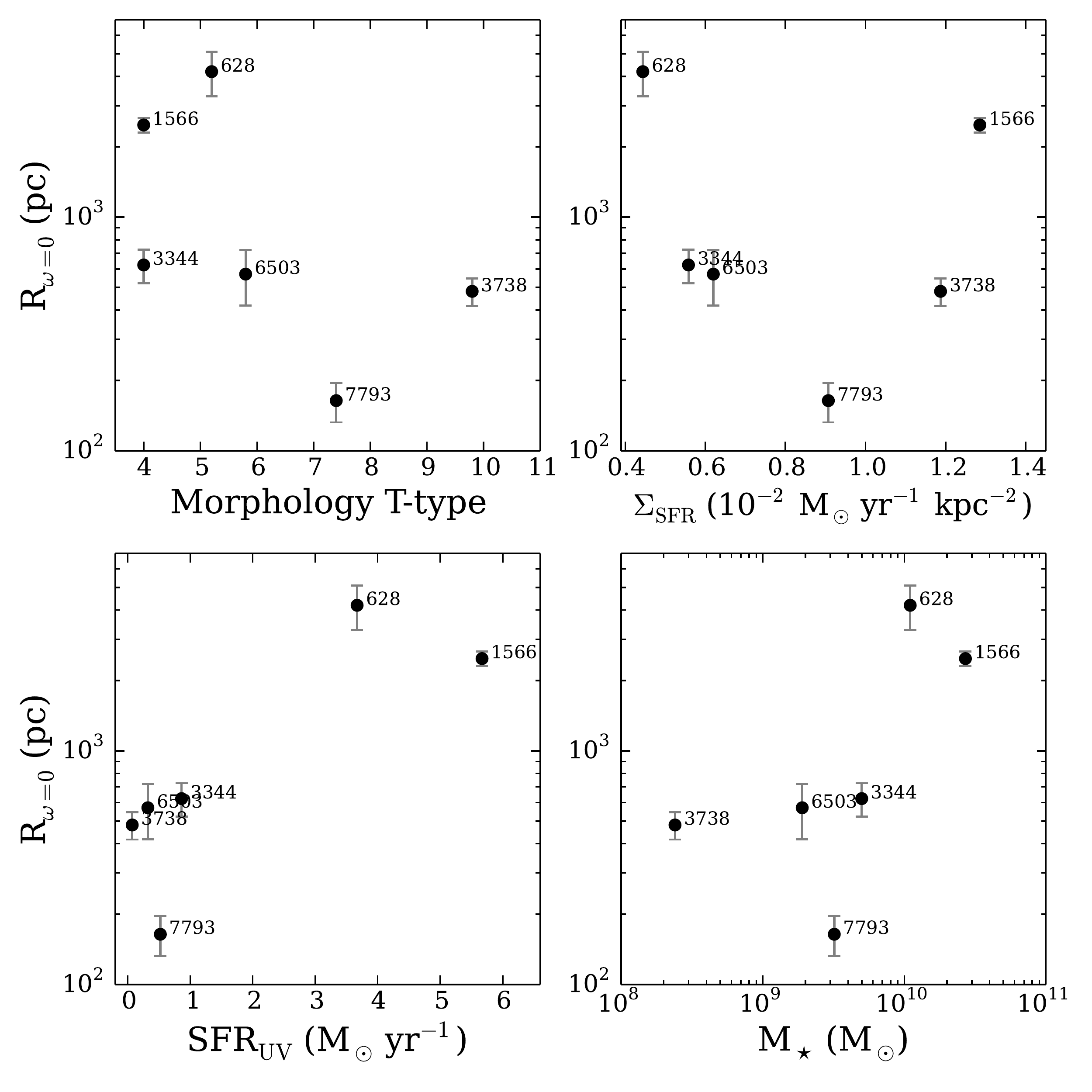}
% made in mag_6gal.py
%BoundingBox:  18 144 575 702
\caption{
Correlation length $R_0$ where the correlation function (Figure \ref{fig:2pcfall}) becomes flat ($\omega=0$) as a function of the morphological type, star formation rate surface density $\Sigma_{\rm SFR}$, the SFR, and the stellar mass of each galaxy.  The galaxy names are shown with their numbers only (without the leading NGC).  There are general trends for larger correlation lengths in galaxies that are larger/brighter, have stronger spirals, and bigger SFRs.  A larger sample of galaxies is necessary to more accurately describe how trends of galactic physical properties influence the clustering results.  No trend is seen between the correlation length $R_0$ and $\Sigma_{\rm SFR}$.  
\label{fig:physprops}}
\end{figure}

Figure \ref{fig:physprops} shows the correlation length of each galaxy compared to different physical properties: morphological T--type, SFR, star formation rate surface density $\Sigma_{\rm SFR}$, and the stellar mass.  In general, larger and brighter galaxies exhibit larger correlation lengths than smaller galaxies.  This implies that the distribution of star-forming regions within larger galaxies are hierarchical over longer spatial scales, both in the strength of the observed clustering and the extent of the complexes ability to survive over larger scales compared to what is observed in smaller galaxies.  Galaxies with higher SFRs and lower morphological T numbers also have a general trend for larger correlation lengths.  This indicates that the young stellar clusters in flocculent and dwarf/irregular galaxies with lower SFRs are distributed more homogeneously at the same spatial scales as observed in spiral galaxies, and thus, these systems globally have smaller star-forming complexes.  There is no trend on the SFR surface density $\Sigma_{\rm SFR}$ of the correlation length, which is surprising as recent observations suggest that the SFR efficiency scales with the SFR surface density \citep{cook16}.  While these observed trends are weak, limited by the small sample sizes, future studies with the full LEGUS sample of galaxies will help improve these trends and investigate why galaxies with larger SFRs correlate with large correlation lengths but no such trend is observed with the $\Sigma_{\rm SFR}$.   %[reason we would expect larger, higher SFR, and more 'evolved spirals' (lower T type) galaxies to have a larger correlation length?]

\subsection{Combined Age Results}\label{sec:combined}
\begin{figure*}
\vspace{-80pt}
%\plotone{2pcf_mg_weighted_all.png}
\includegraphics[scale=.65,bb=0 0 1008 720]{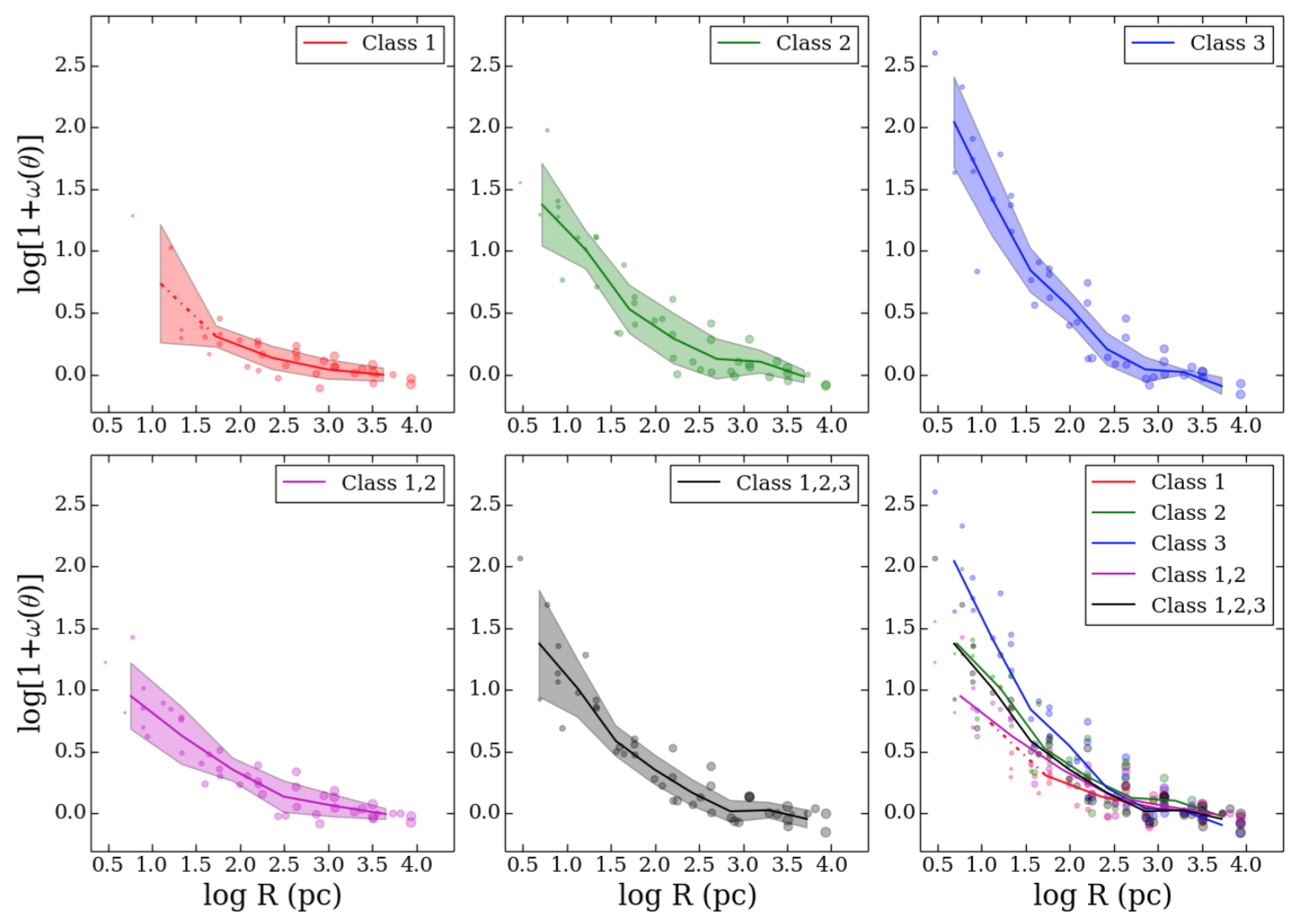}
% made in /weighted/weighted_python.py
%BoundingBox:  18 144 575 702
\caption{
The weighted two-point correlation functions $1+\omega(\theta)$ for all six galaxies as a function of physical scale (parsec) separated by cluster class.  The solid line shows the weighted average of each class, weighted by number of pairs in each data point.  The size of the data points represents the number of pairs for each point and the shaded region represents the $1\sigma$ scatter.  We represent power-law fit for the class 1 clusters at short separations with a dashed line as we are limited by very small numbers within that bin for those clusters.   
\label{fig:2pcfall_w}}
\end{figure*}

In order to compare the results and analyze how all clusters behave globally across all galaxies, we combine each cluster class across all galaxies by summing the weighted number of pairs as a function of physical separation, as shown in Figure \ref{fig:2pcfall_w}.  The improved statistics from the increase in clusters per class permits to better investigate common age effects among the galaxies.  We fit a power law to each cluster class to the total sample of clusters and to the clusters above/below an age of 40 Myr, where that choice in age comes from the typical median age where each galaxy shows maximal clustering (see Figure \ref{fig:2pcfage}).  Table \ref{tab:4} shows the power law fits for the weighted average correlation for each class.  Up to a breakpoint of 112 parsecs, the entire cluster population {\it averaged over all six galaxies} has a recovered power law slope of $\alpha =-0.83$, very similar to the fractal dimension of both stars and local interstellar clouds \citep[see, e.g., ][]{falgarone91,elmegreen06,sanchez08}, inherit their hierarchy at birth and preserve the natal gas pattern for a few tens of Myrs.  The global slope for all clusters is very similar to the slope recovered for the class 2 clusters of $\alpha =-0.85$, whereas the combined 1+2 clusters have a shallower slope of $\alpha =-0.57$ and class 1 clusters have a very shallow, single slope of $\alpha =-0.18$ across all spatial scales, despite the steep upturn at the smallest spatial scales, which arises from small numbers within that bin (see Figure \ref{fig:2pcfall_w} for the scatter).  Associations, on the other hand, exhibit a much steeper slope with $\alpha =-1.12$.  

As can also be seen in Figure \ref{fig:2pcfall_w}, every classification is observed to display a broken power law, which is not necessarily true for each individual galaxy (Figure \ref{fig:2pcf}).  The reason behind this is that we are increasing statistics by combining clusters from numerous galaxies, which enables us to recover more subtle effects.  This is best described in Figure \ref{fig:2pcfall}, where, for example, the correlation function for NGC 6503 and NGC 3738 becomes flat at a smaller spatial scale than what is observed for NGC 628, likely a difference that arises from the global galactic properties (see \ref{fig:physprops}).  As a result, when the correlation function for all the galaxies are averaged together, there is a general break around 100-200 pc for all cluster types, where the slope beyond the break for all cluster types is nearly uniform ($\alpha \sim -0.2$) and only varies at the smallest spatial scales before the breakpoint.  

\begin{deluxetable}{lccccc}
\tabletypesize{\scriptsize}
\tablecaption{Power-Law Parameters for Weighted Average\label{tab:4}} 
\tablecolumns{6}
\tablewidth{0pt}
\tablehead{
\colhead{Class}				& 
\colhead{$A_1$}				&  
\colhead{$\alpha_{1}$} 	&
\colhead{$\beta$}   		&
\colhead{$A_2$}				&  
\colhead{$\alpha_2$} 	
\\
\colhead{}		& 
\colhead{}		& 
\colhead{} 		&
\colhead{(pc)}   &
\colhead{} 		&
\colhead{}   
}
\startdata  
\hline \noalign{\vskip 2mm}   
\multicolumn{6}{c}{All Ages}\\
\hline \noalign{\vskip 1mm}   
Class 1 					& 62(2)	&  $-0.86(2)$   & 59 		& 3.2(3) & $-$0.14(3) \\
Class 2 					& 63(3)	&  $-0.73(7)$ &  112		& 4.8(3) 	  & $-$0.18(3) \\
Class 3 					& 463(40)&  $-1.13(7)$ &  	112		&	6.6(4) 	  & $-$0.23(07)\\
Class 1,2 				& 25(4)	&  $-0.58(4)$ &  93		& 3.6(2) 	  & $-$0.15(2) \\
Class 1,2,3 				& 69(5)	& 	$-0.77(7)$& 	112		&	4.0(3) 	  & $-$0.17(4) \\
\hline \noalign{\vskip 2mm}   
\multicolumn{6}{c}{Age $\leq 40$ Myr}\\
\hline \noalign{\vskip 1mm}        
Class 1 					& 52(10)		&  $-0.66(14)$		& 128 		&	6.4(3) 	& $-$0.23(3) \\
Class 2 					& 119(15)	&  $-0.83(3)$		&  112		& 8.1(5)	& $-$0.26(2) \\
Class 3 					& 565(18)	&  $-1.15(13)$		&  112		&	7.3(4) 	& $-$0.232(16) \\
Class 1,2 				& 39(2)		&  $-0.60(3)$ 		& 186		& 3.51(13) & $-$0.139(19) \\
Class 1,2,3 				& 115(6)		& 	$-0.85(3)$ 		& 	112		&	5.6(3) 	& $-$0.21(2) \\
\hline \noalign{\vskip 2mm}   
\multicolumn{6}{c}{Age $> 40$ Myr}\\ 
\hline \noalign{\vskip 1mm}   
Class 1 					& 2.45(14)	&  $-0.13(2)$ 		& \multicolumn{3}{c}{\nodata}\\
Class 2 					& 3.98(10)	&  $-0.160(18)$ 	& \multicolumn{3}{c}{\nodata}\\
Class 3 					& 3.3(3)		&  $-0.161(14)$ 	& \multicolumn{3}{c}{\nodata}\\
Class 1,2 				& 2.90(13)	&  $-0.130(18)$ 	& \multicolumn{3}{c}{\nodata}\\
Class 1,2,3 				& 2.61(15)	&	 $-0.164(6)$ 	& \multicolumn{3}{c}{\nodata}
\enddata
\tablecomments{
Columns list the 
(1) Classification of stellar clusters; 
(2) Amplitude $A_1$ of the angular correlation function before the breakpoint; 
(3) Slope $\alpha_1$ of the angular correlation function after the breakpoint; 
(4) Location of the break point $\beta$ in parsecs; 
(5) Amplitude $A_2$ of the angular correlation function past the breakpoint; and
(6) Slope $\alpha_2$ of the angular correlation function after the breakpoint.  All values are derived by using as independent coordinate R in parsecs.  Numbers in parentheses indicate uncertainties in the final digit(s) of listed quantities, when available. 
}
\end{deluxetable}
-0 -0 576 432
\begin{figure*}
%\epsscale{1.}
%\plotone{2pcf_mg_weighted_all_agecut.png}
\vspace{20pt}
\includegraphics[scale=.47,bb=0 0 1008 720]{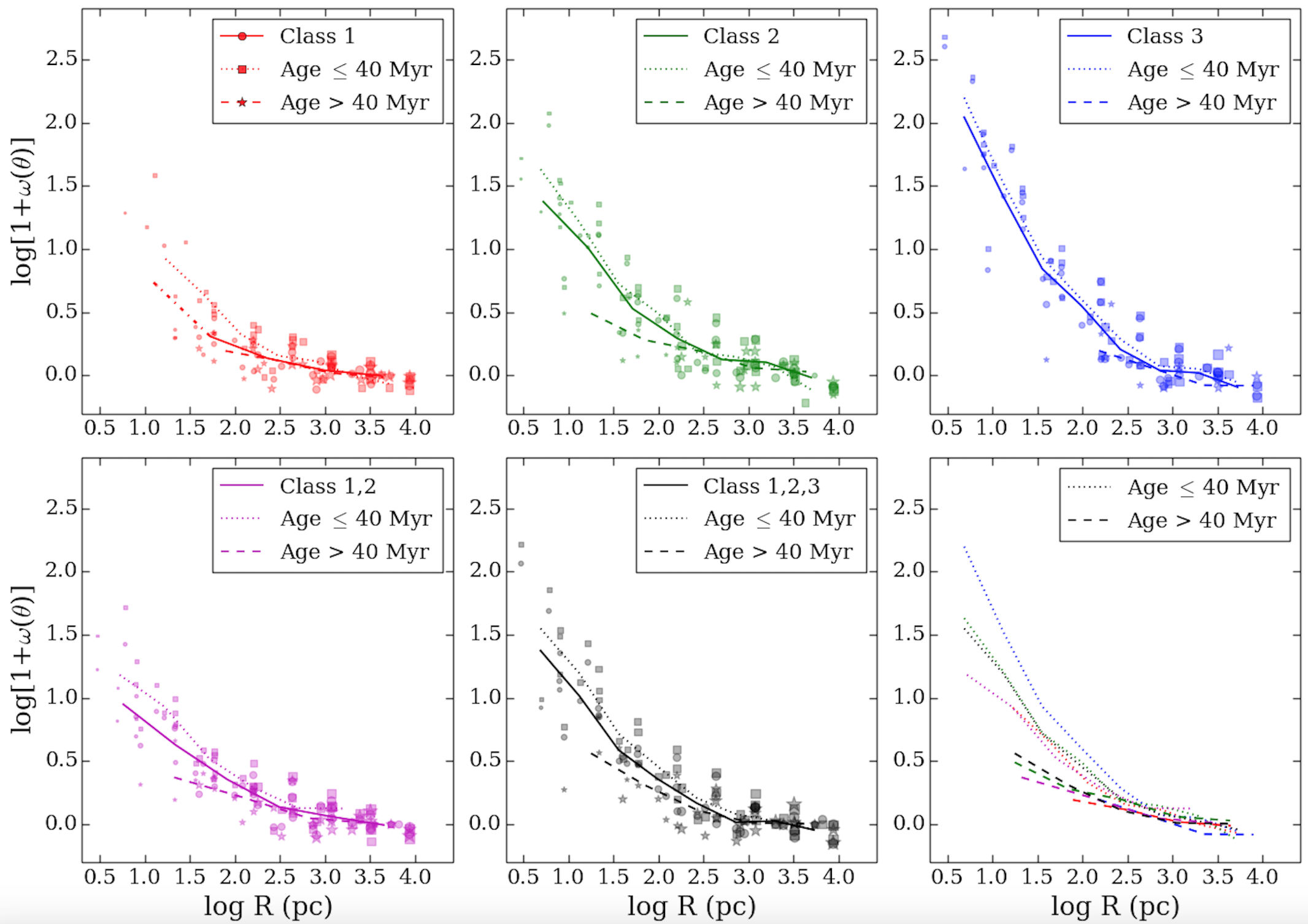}
%/weighted/weighted_python_agecut.py
%BoundingBox:  18 144 575 702
\caption{
The weighted two-point correlation functions $1+\omega(\theta)$ as a function of physical scale (pc) divided by age for each cluster class.  The solid line (and circle symbols) show the weighted average of each class, weighted by number of pairs in each data point, the dotted line (and square symbols) shows the weighted average for clusters with ages less than 40 Myr and the dashed line (and star symbols) represents clusters older than 40 Myr.  The size of the data points represents the number of pairs for each point.  We represent the class 1 clusters at short separations with a dashed line as we are limited by very small numbers within that bin for those clusters.  The last panel shows all the trends for clusters older/younger than 40 Myr. 
\label{fig:2pcfall_w_age}}
\end{figure*}

Figure \ref{fig:2pcfall_w_age} shows how the ages of clusters influence the global weighted average correlation function across all six galactic environments.  We divide the clusters into their class and further separate into age bins of older/younger than 40 Myr and compute the correlation function for each trial, taking the weighted average across all galaxies.  As can be seen, for each class of cluster, the strength of the clustering disappears for clusters of older ages.  While there is large scatter between all the galaxies, by ages of 40 Myr, the clustering strength has already decreased across all systems.  

What is significant is that the behavior of the class 1 and 2 clusters older than 40 Myr show an appreciable flattened distribution out to small spatial scales ($\sim$10 pc), indicating that the clusters that survive past 40 Myr become randomized at all spatial scales.  The class 3 associations, on the other hand, show an absence of older ($\gtrsim$40~Myr) clusters within $\lesssim$100~pc of each other, suggesting early dissolution.  We conclude that the observed decrease in clustering strength is an effect of the cluster classes marking an age sequence in the cluster classes.  More importantly, this common behavior is observed in six different galaxies characterized by quite different environments, as defined by their morphologies, which ranges from Sbc to Irregular, indicating that the clusters randomize on fairly short timescales.

\subsection{Binary Clusters}\label{sec:binary}
The hierarchical structure of star formation results in the highest efficiency of star formation occurring within dense, crowded environments at the smallest scales.  This has the possibility in resulting in binary or groupings of multiple star clusters at small scale lengths that are capable of interacting with other.  The existence of binary star clusters have been observed in the LMC \citep{bhatia88,dieball02} and the MW \citep{subramaniam95,delafuentemarcos09}.  

Most cluster pairs are observed to have projected separations less than 20 pc, relatively young ages, and show small age differences between the pairs, indicative that the pair was formed at birth.  Thus, binary star clusters are believed to be short--lived, with lifetimes of 10--100 Myr \citep{bhatia90,delafuentemarcos10,priyatikanto17}.  Within the Magellanic Clouds, between $\sim$10--20\% of clusters are potentially part of a binary/multiple system \citep[see][and references therein]{dieball02}.  The MW shows a lower prevalence of binary star clusters, however, this could be an observational effect, resulting from the relatively short lifetime of binary star clusters compounded by the paucity of young star clusters observed within the MW \citep{portegieszwart10}.  

Assuming a maximum cluster pair separation of 20 pc is required for a binary system to be bound and interacting, this separation is smaller than the closest cluster pairs we observe at ages older than 40 Myr  (Figure \ref{fig:2pcfall_w_age}, dashed lines in the bottom right plot).  We conclude that none of our clusters reside in bound binaries or groupings at these separations past this timescale.  

When we consider clusters with ages less than 40 Myr, about 6.5\% of our total cluster sample (241 clusters out of 3685 total across all six galaxies) includes pairs closer than 20 pc, and could be binary clusters.  Recent results of young star clusters in the same LEGUS galaxies show that pairs of clusters with small separations are more likely to be coeval and exhibit similar ages \citep{grasha17}.  The same findings of similar ages with small separations for pairs of clusters is also observed in the MW \citep{efremov98} and the LMC \citep{delafuentemarcos09}.  Similar ages for small separations is an expected result from turbulence driving the hierarchical structure of the ISM \citep{elmegreen96}.  These complementary studies indicate that clusters favor being born close together at roughly the same time within groupings of multiple star clusters, resulting in the observed strong clustering at the youngest ages and the smallest separations.  The clustered complexes disperse with age, effectively decreasing the observed clustering and removing the cluster pairs at the shortest separations (Figure \ref{fig:2pcfall_w_age}).  However, we are unable to dynamically constrain if close cluster pairs are gravitationally bound to each other, or if they will simply disperse with age.

%At the distances of the LEGUS galaxies ($\sim$3.5--15 Mpc), star clusters cannot be resolved into their individual stars; they appear as being slightly more extended than the stellar point spread function.  The standard technique for deriving cluster properties -- masses and ages -- is through fitting the observed integrated light to computed predictions of the clusters SEDs \citep[see Section \ref{sec:clusterselection} and][]{adamo17}.  The best-fit correspond to the value associated with the best (i.e., smallest) $\chi^2$ and the quality of the fit is evaluated with the reduced $\chi^2$ value across all the bands.  We also list the Q-probability as a measurement of the quality of the fit: Q=1 suggests a fit that is consistent with the data and Q=0 suggests a poor fit.  

%Chance alignment of stars or star clusters will impact our class 3 sources more than class 1 or 2 clusters, which in general are more isolated.  In general, the fits to the class 3 clusters for NGC 628 are more unconstrained: class 1 and 2 clusters have a reduced $\chi^2$ value of 1.0 and a Q-probability of 0.38, whereas class 3 associations have have a reduced $\chi^2$ value of 1.3 and a Q-probability of 0.29.  

\section{Discussion}\label{sec:discussion}
In this work we have identified stellar clusters and associations to investigate, their clustering properties with respect to the rest of the stellar clusters in each galaxy, and the timescale over which the clustered substructure that these clusters reside in dissipates.  For a hierarchical star formation model, the turbulent-driven ISM induces scale-free star formation across all spatial scales, resulting in the clustering of star clusters and creating substructure on pc-scale scale lengths, dispersing and distributing over wider areas as they age \citep{portegieszwart10}.  Amid this general framework \citep[e.g., ][]{zhang01,gouliermis15}, there are still many unanswered questions relating the properties of star clusters to the process of star formation, including (1) the timescale for the survival of the substructure; (2) the extent that the environment influences cluster formation, evolution, and disruption, and hence, these hierarchical star-forming complexes; and (3) how gravitationally bound stellar systems behave compared to loosely bound stellar systems (associations), as dense cloud systems and associations may differ as early as birth \citep[see, e.g., ][]{wright14}.  

%Early observations \citep{grasha15} of $\sim$1200 young stellar clusters within NGC 628 using the two-point correlation function found the young star clusters exhibited a clustered, self-similar distribution and that the randomization timescale to become uncorrelated relative to each other is around 40 Myr.  

The two-point correlation function provides a means to investigate and quantify the stellar aggregate distribution within extended star-forming structures and accurately determine the randomization timescale for the components to become uncorrelated relative to each other.  We find that the young clusters are clustered with respect to each other and that the distribution randomizes with increasing age.  Despite spanning different environments, with morphological types from Sbc to irregular, similar dispersion timescales of a few tens of Myr are observed in all six galaxies.  

The visual identification technique we implement allows us to distinguish between class 1 and 2 clusters, which are potentially gravitationally bound stellar systems, and class 3 associations, providing us the ability to investigate how these two different types of stellar aggregates evolve over time.  The young age distribution and the morphology, characterized by multi-peaks and asymmetries in the light distribution, indicates that these clusters are most likely stellar associations, unbound stellar systems that evaporate and disperse on short time scales \citep[tens of Myr;][]{gielesportegieszwart11}.  This is additionally supported by \citet{adamo17} for star clusters in NGC 628, finding that class 3 associations tend to be systematically younger and to have a steeper mass function than class 1 and 2 clusters.  We find that associations behave differently in their clustering properties compared to the class 1 and 2 compact clusters across all galaxies.  Class 1 systems have relatively weak clustering and in general are the most massive cluster type and exhibit older ages.  At the other end of the spectrum, class 3 associations display a very young age distribution and exhibit very strong clustering within each galactic system.  In addition, class 3 associations display a steep decrease in their numbers with increasing age and exhibit a median age that is always below 10~Myr (Figure \ref{fig:hist}).  The short timescales of survivability further reinforce the idea that these stellar systems are not likely gravitationally bound systems and they disrupt soon after formation, and thus, are still located near their birth sites, in agreement with the observed increase of clustering for class 3 sources.  For a cluster of mass $10^4$~M$_{\odot}$ and a radius of 5 pc, the crossing timescale within that cluster is a few Myr.  Hence, for class 1 and 2 clusters with an average population age $\gtrsim 10$~Myr, the crossing time is shorter than the age, and we can consider these clusters as potentially gravitationally bound systems.  
%Infant mortality occurs within the first 7-10 Myr of the life of a cluster. These associations live longer, a few tens of Myr, so perhaps are not `victims' of infant mortality?
% M=10^4 Msolar, R= 5pc,
% v=(0.4*G*M/R)**1/2 == 1.86 km/s
% t_cr = R/v = 8.3*10**13 s = 2.6 Myr

Recent work on the number densities of the young stellar clusters in NGC 628 by \citet{adamo17} also shows a rapid decline in class 3 associations, disappearing on time scales of $\sim$50 Myr, comparable to those of hierarchically structured star-forming regions that we find in this study.  Furthermore, class 1 clusters appear to already be less clustered at birth compared to the class 3 associations \citep{wright14,adamo17}, as seen in Figure \ref{fig:2pcfall_w_age}: when we only consider clusters less than 40 Myr, class 1 clusters still display a flatter clustered distribution compared to what is observed for class 3 associations.  Thus, class 1 clusters do not trace the multi-scale hierarchy as directly as the associations, which may reflect the distribution of the densest peak (the putative birth sites of the bound clusters) within the ISM hierarchical structure.  The increased clustering observed with young clusters -- of all morphologies, though the clustering strength is strongest for the associations -- supports the idea that all recently born stars and star clusters are formed within hierarchical star-forming complexes and their distribution reflects the fractal patterns of their parent molecular clouds.  The relatively quick dispersal time of the associations could provide a significant contribution to the stellar field population of their host galaxies \citep[see, e.g., ][]{maizapellainiz01}.  

Studies to constrain the properties of young star clusters, connecting them with their surrounding environment, are reasonably straightforward and achievable with extragalactic studies.  It is more difficult to conduct studies of young star clusters within the disk of the MW galaxy, as the observations face complications such as foreground/background confusion, distance uncertainties, and severe line-of-sight reddening, which makes it arduous to detect young stellar systems.  Despite these challenges in detecting and interpreting the observations of stellar clusters in the MW galaxy, the young stellar clusters within 2~kpc of our Solar System exhibit evidence of formation within hierarchical structures from the observed correlation of their age and separations, possibly driven by a turbulent ISM \citep{delafuentemarcos09}.  The same observational signature is also present in the age and positions of the young star clusters in the LMC \citep{efremov98} and other local galaxies \citep{grasha17}.  Additionally, local star-forming environments in the MW display distributions that are consistent with predictions of hierarchically structured star-formation \citep{bressert10} that is also observed in extragalactic studies, arising from the molecular cloud hierarchical structure \citep{elmegreen08,elmegreen14}. 

This study finds that the youngest clusters ($<$50 Myr) primarily trace the spiral arms while older clusters are more randomly distributed throughout the galaxy.  We conclude that star-forming complexes are relatively young, predominately located within the spiral arms of their galaxies, and are among the largest scales of the star formation hierarchy.  The total amount of global star formation captured within clustered structures versus the amount of star formation that is `unclustered' (i.e., the stellar field population) has important consequences on the mode (i.e., one universal mode or multiple, distinct modes) of star formation.  We will investigate and address these issues in a forthcoming paper.  

If star clusters have a velocity dispersion of $\sim$1~\kms\ \citep{larson81} typical of their birth clouds, the clusters are capable of traveling 1 parsec per Myr.  Shear effects, spiral motions, and colliding GMCs can compound the initial random motion, capable of boosting the velocity up to 10 \kms\ \citep[see, e.g., ][]{whitmore05,gustafsson16}.  Stellar systems with a total random motion of 10~\kms\ can travel 100 pc within 10~Myr, making a 50~Myr timescale for randomization of star clusters within star-forming substructures entirely reasonable, resulting in a correlation length of a couple hundred of parsecs over that time period consistent with our findings.  

The amplitude in the correlation function reflects the difference in structure between the galaxies, the amplitude scales with the magnitude of the host-galaxy: for a fixed scale, the faintest galaxy, NGC 3738, has the smallest value of $\omega+1$, and the brightest galaxy, NGC 1566, has the largest value (Figure \ref{fig:2pcfall}).  The different slopes and clustering strengths exhibited by the different classes and ages of clusters imply different fractal dimensions and hierarchies for different galaxies, suggesting that each galaxy has its own influence in the formation and survival of star-forming structures.  Despite the significant variations in the power-law slope -- a proxy for the fractal dimension -- among the galaxies, a global average over all galaxies with $\alpha =-0.83$ is consistent with the stellar clusters forming from fractal gas in a hierarchical star formation framework.  While we have found that the local environment does influence the distribution of the stellar systems -- larger galaxies with higher SFRs have larger clustering correlation lengths -- the exact influence of the local environment on observed differences in the global clustering results and on the nature of the 20--60 Myr randomization timescale -- requires a larger sample of galaxies that span a broad range of physical parameters.

\section{Summary and Conclusion}\label{sec:summary}
We present a study of the clustering of the young stellar clusters in six nearby galaxies, drawn from the LEGUS sample of 50 local star-forming galaxies with UV and optical data taken with WFC3/UVIS from HST, with the addition of archival optical ACS images.  Taking advantage of both the high-angular resolution observations and the reliable measurements of ages and masses, we identify and visually classify the brightest star clusters (luminosities in the V-band brighter than $-6$ mag).  All stellar clusters are divided into two broad categories of morphology:  compact cluster candidates -- according to our classification scheme, we consider these to be bound cluster candidates -- and multi-peak systems, likely expanding or loosely bound stellar associations.  In total, we identify 2323 compact clusters and 1362 associations, resulting in a total of 3685 stellar clusters in six galaxies of varying distance from 3.44 Mpc to 13.2 Mpc.  Multiband cluster photometry is combined using deterministic single stellar population models to derive an age, mass, and color excess E(B--V) for every star cluster with errors of $0.1$~dex or less for all derived measurements. 

We implement the angular two-point correlation function to quantify the clustering of star clusters within each galaxy, finding that the correlation functions are well described with a single power-law slope, and in some cases, a double power-law.  The general observed correlation between young stellar clusters decreases monotonically as a function of separation, as expected in hierarchical structures.  We also find that the clustering strength is stronger for the younger clusters compared to the older ones at the same separation lengths, which tend to be weakly or non-clustered within each galaxy and across all cluster types.  Generally, the mass of the star clusters has little effect on the clustering strength; it is primarily governed by age.  The near-flat slopes measured for the older clusters corroborate the effectiveness of the two-point correlation function to differentiate between a randomly distributed population from that of a clustered, fractal distribution. 

We conclude that the observed decrease in clustering strength is the result of the clusters taking on more uniform positions throughout their galaxy, erasing the observed substructure with time.  The timescale for the clustering erasure for the hierarchical clustering begins around 20--60 Myr, and the characteristic size of the clustered structures is about 100--300 pc across different galaxies.  Thus, these identified star-forming complexes are very young.  

The morphological classifications -- compact clusters and stellar associations -- not only provide robust cluster catalogs free of stellar contaminants, but also yield insight into the properties of the stellar systems.  While both the stellar clusters and associations inherit the imprints of the hierarchical structure of the ISM at birth, their evolutionary paths differ from each other after their formation.  The compact stellar clusters survive for longer time periods, with median ages of 9--40 Myr \citep[and their ages are likely to become older, but we are limited by selection effects, see][]{adamo17}, whereas associations display very young median ages between 2--10~Myr.  The difference in the physical properties of the bound stellar systems compared to the associations has an important impact on the clustering properties of the two stellar systems: associations are very strongly clustered in all six galaxies and in comparison, the (older) compact stellar clusters are less strongly correlated in their spatial distributions.  We conclude that clusters display a more homogeneous distribution as a result of their birth sites being more homogeneously distributed.  

Finally, we take the weighted average correlation of all the star clusters across all six galaxies as a function of spatial scale to investigate the average global clustering observed across the different environments.  We fit a power-law to the weighted average correlation function of all the clusters in all the galaxies and measure a slope of $-0.77$, expected for a star formation process dominated by turbulence.  The age-dependency to the clustering of star clusters found for individual galaxies becomes notably compelling when averaged across all galaxies: for all clusters with ages less than 40 Myr, we find that the clustering slope is flat and uniform, with values between $-0.13$ and $-0.16$, regardless of cluster type.  This indicates that star clusters  rapidly disperse from their hierarchically organized star-forming regions  regardless of the galaxy type or size of the region.  There are a large range of slopes for each individual galaxy, especially observed in the distribution of the youngest (age $\lesssim$40 Myr) clusters, indicative that there is a range of fractal hierarchies between galaxies and that the local environment can influence the initial formation of star clusters within the clustered structures.  Indeed, initial results do suggest that larger correlation lengths and stronger clustering is exhibited in galaxies that are larger and have higher SFRs.  Identifying the exact physical nature for the randomization timescale -- and how the local and/or global environment, such as the difference in timescales for spirals versus ellipticals or the inner versus outer regions of a galaxy, influences the results -- is an important goal for future work when a larger sample of galaxies becomes available.  

%, a natural step forward for this study is to understand how the stellar assembling process relates to the structuring behaviour of the natal ISM. This topic will be addressed in a subsequent study, where we compare the interstellar gas structure to that presented here 

%Our main results can be summarized as follows:
%\begin{enumerate}
%	\item item 1
%	\item item 2
%	\item item 3
%	\item item 4
%\end{enumerate}

\acknowledgements
We thank the referee for their careful reading of this manuscript and providing comments that have improved the scientific outcome and quality of the paper.
AA thanks the Swedish Royal Society (KVA) for the awarded founding.  MF acknowledges support by the Science and Technology Facilities Council [grant number ST/L00075X/1].  DAG kindly acknowledges financial support by the German Research Foundation (DFG) through program GO\,1659/3-2.
Based on observations made with the NASA/ESA Hubble Space Telescope, obtained at the Space Telescope Science Institute, which is operated by the Association of Universities for Research in Astronomy, under NASA Contract NAS 5--26555.  These observations are associated with Program 13364 (LEGUS).  Support for Program 13364 was provided by NASA through a grant from the Space Telescope Science Institute.  
This research has made use of the NASA/IPAC Extragalactic Database (NED) which is operated by the Jet Propulsion Laboratory, California Institute of Technology, under contract with NASA.

\appendix 
\section{The Effect of Different Dust Models on the Clustering Results}\label{sec:appA}
We use the photometric catalogs to derive the ages of our star clusters using a SED fitting algorithm, which we will touch upon below, but are available in detail in \citet{adamo17}.  The clusters in this paper for their SED fits assume a single stellar population (SSP) using two stellar libraries to create two sets of SSP models, Padova AGB and Geneva tracks without rotation, both available in Starburst99 \citep{leitherer99}.  We assume the initial mass function (IMF) is fully sampled and adopt a Kroupa IMF, with stellar masses between 0.1 and 100 M$_{\odot}$.  The models are reddened prior to SED fitting the photometry and the grid includes increasing internal reddening of E(B--V) = [0.0, 1.5] with steps of 0.01 mag.  The fit incorporates three different extinction and/or attenuation laws: (1) the Milky Way extinction law from \citet{cardelli89}; (2) the starburst attenuation law by \citet{calzetti00}, assuming the stars and gas suffer the same reddening; and (3) the differential starburst attenuation law where we assume the gas emission suffers higher extinction than the stars. 

In total, there are 12 star cluster catalogs available for each galaxy, produced with deterministic models and fitting procedures: a combination of the two photometric approaches for aperture correction \citep[average based and CI based; see][for a detailed analysis and treatment on how these two approaches affect the derived properties]{adamo17}, two stellar libraries (Geneva and Padova AGB), and three extinction/attenuation curves (Milky Way, Starburst, and differential Starburst).  For a few galaxies, there are also catalogs based on using a Bayesian analysis method together with stochastically sampled cluster evolutionary models, presented by \citet{krumholz15} using the Stochastically Lighting Up Galaxies \citep[SLUG; ][]{dasilva12} code.  We do not incorporate the stochastic models in our results and refer the reader to \citet{krumholz15} for a detailed analysis on how the deterministic approach affects the derived cluster properties compared to the stochastic approach.  

Of the 12 deterministic models available, our reference cluster catalog uses the Padova stellar evolutionary models assuming solar metallicity and the starburst attenuation law.  As different flavors of stellar libraries and assumptions in the dust geometry used to build cluster evolutionary tracks are bound to influence the derived cluster properties, we check here how our results depend on the assumptions made to derive the physical properties from the cluster photometry.  Because our primary science goal is determining the timescale for which the hierarchical cluster structures randomize and dissipate, happening on scales of a few tens of Myr, we examine the effect that the different models will have on the derived age of our stellar clusters.  

\begin{figure*}
%\epsscale{1.15}
\includegraphics[scale=.42]{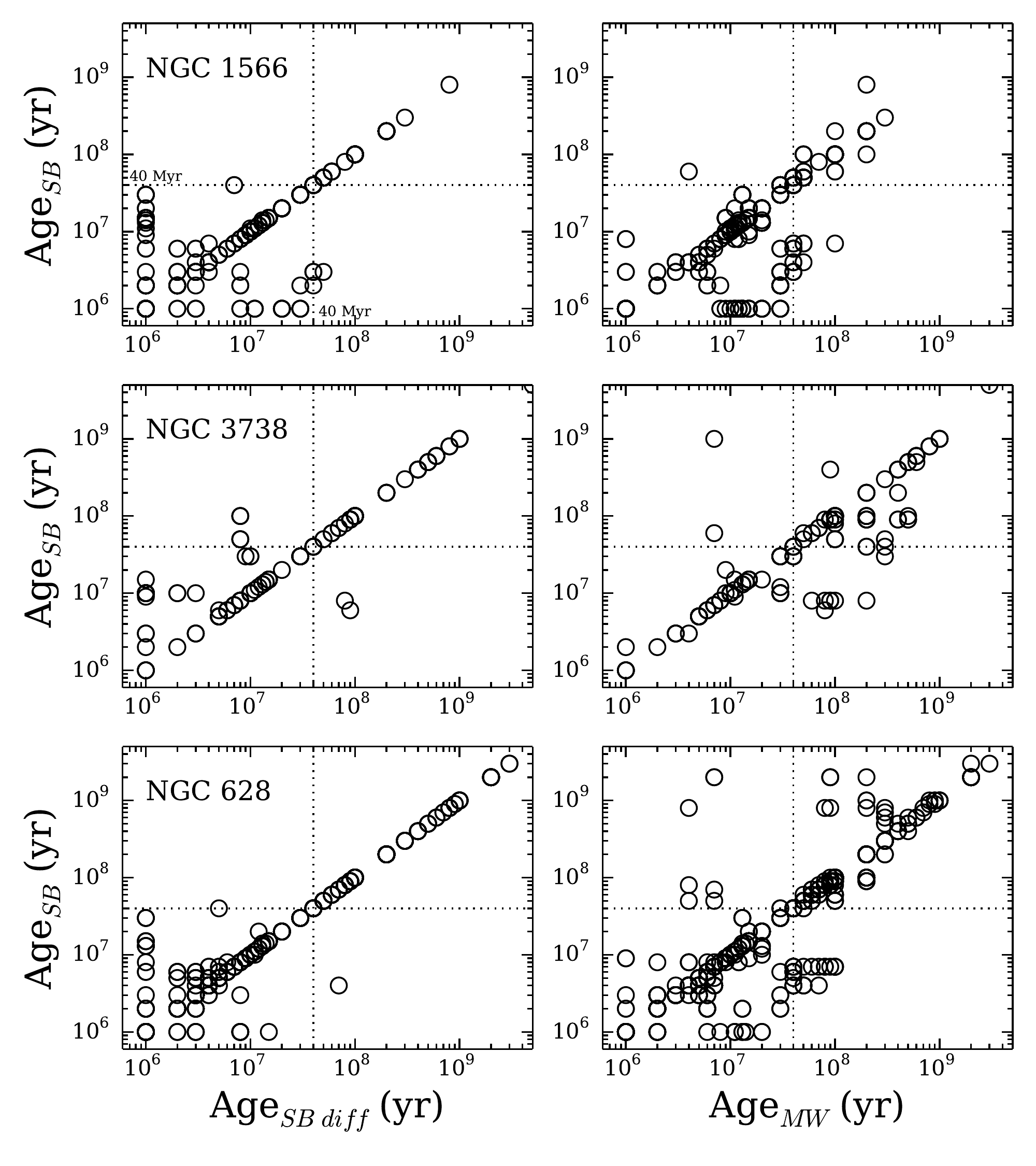}
\includegraphics[scale=.42]{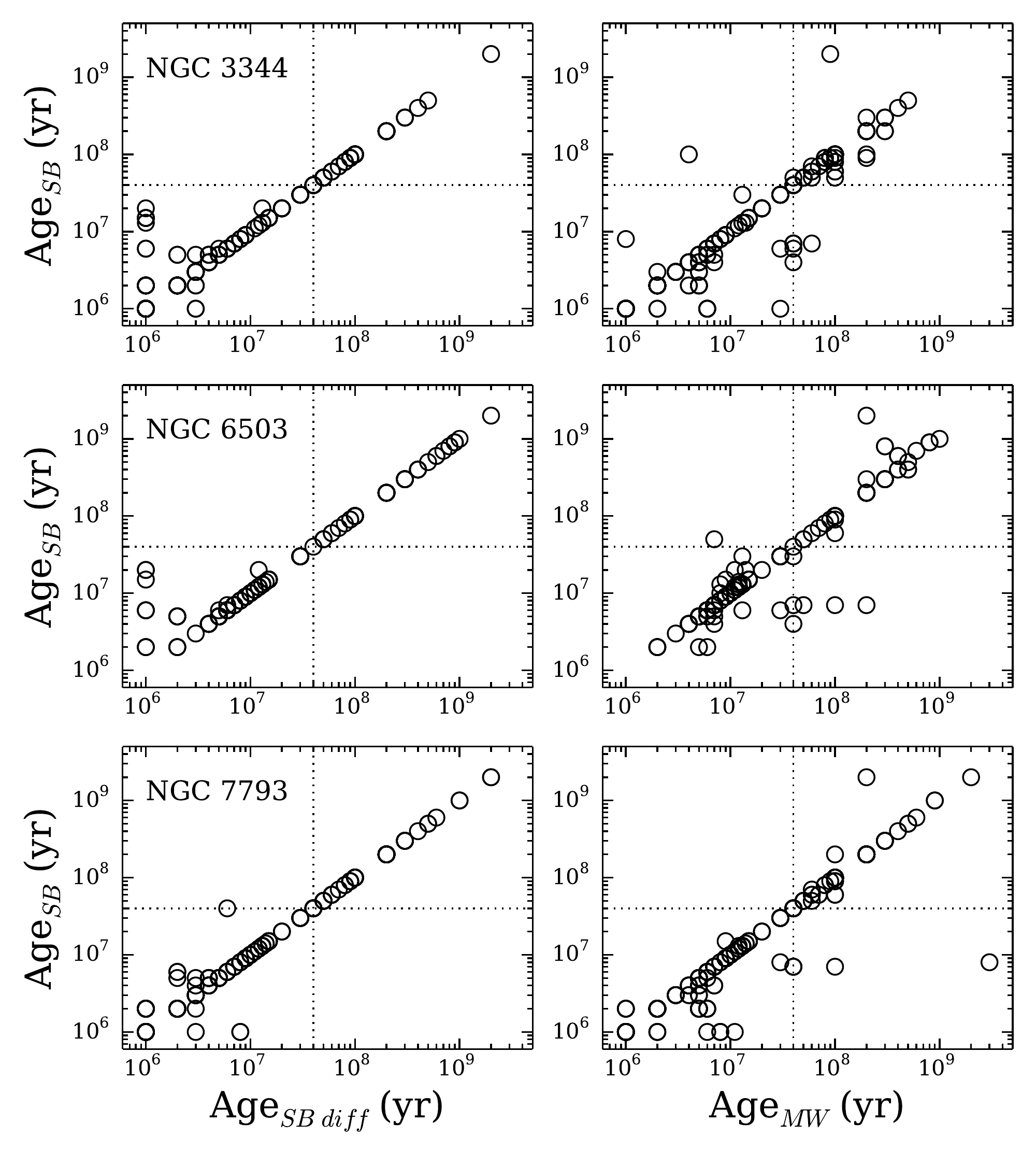}
%\plottwo{compare_extinction_1.eps}{compare_extinction_2.eps}
% /Research/Thesis/legus_clusters_v13/compare_extinction.py
\caption{
The age of the star clusters, as determined using a starburst (SB) attenuation curve, as a function of the age as determined with a differential starburst attenuation curve (SB diff; left column) and a Milky Way extinction curve (MW; right column), for all six galaxies in the sample.  The horizontal and dotted lines delineate an age cutoff of 40 Myr.  The one-to-one correlation in the ages deviates most at the youngest ages (primarily below $\sim 10$~Myr), which is expected, as the shortest wavelengths are where the different dust models have the greatest deviation from each another.  
\label{fig:compareext}}
\end{figure*}

The two-point correlation functions of Figure \ref{fig:2pcf}, \ref{fig:2pcfall}, and Figure \ref{fig:2pcfall_w} are independent of derived cluster properties as they only take into account the location of each cluster within the host galaxy.  The cluster properties only start to influence the results when we make bins in age, looking at the clustering as a function of both spatial scale and age, as we do in Figure \ref{fig:2pcfage} and \ref{fig:2pcfall_w_age}.  Figure \ref{fig:compareext} shows the age of the clusters within each galaxy, derived using our reference starburst attenuation model, as a function of age of the clusters determined using both the differential starburst and Milky Way extinction law.  We also show the line that delineates where we make our age bin for clusters older/younger than 40~Myr for Figures \ref{fig:2pcfage} and \ref{fig:2pcfall_w_age}.  Ideally, we want to avoid models where the clusters appear in the upper left hand or lower right hand part of each plot, as these clusters will move between the old/young age bin depending on which dust model is assumed.  The starburst and differential starburst laws show fairly consistent one-to-one relations in the recovered ages, where the greatest deviations are at the youngest ages, primarily confined to ages below $\sim 10$~Myr.  The youngest ages will have the biggest flux contribution from the shortest wavelengths (i.e., UV) and subject to the larger effects from attenuation, responsible for the increase in age spreads at younger ages.  The starburst and Milky Way derived-ages deviate more relative to each other, and across a larger age range, especially within NGC 628 and NGC 3738.  

To better understand the exact influence of the different models on the derived properties, we recompute the two-point correlation function analysis for NGC 628 using the Milky Way extinction curve and NGC 3738 using the differential starburst curve attenuation curve, and compare them to the results in Figure \ref{fig:2pcfage} using the starburst model curve.  A large number of sources moving between the age-separation at 40 Myr will have a substantial effect on the clustering results.  

\begin{figure}
%\epsscale{1.15}
%\plotone{ngc3738_deproj_2pcf_agecut_4panel_diffSB_class123.eps}
\includegraphics[scale=.35]{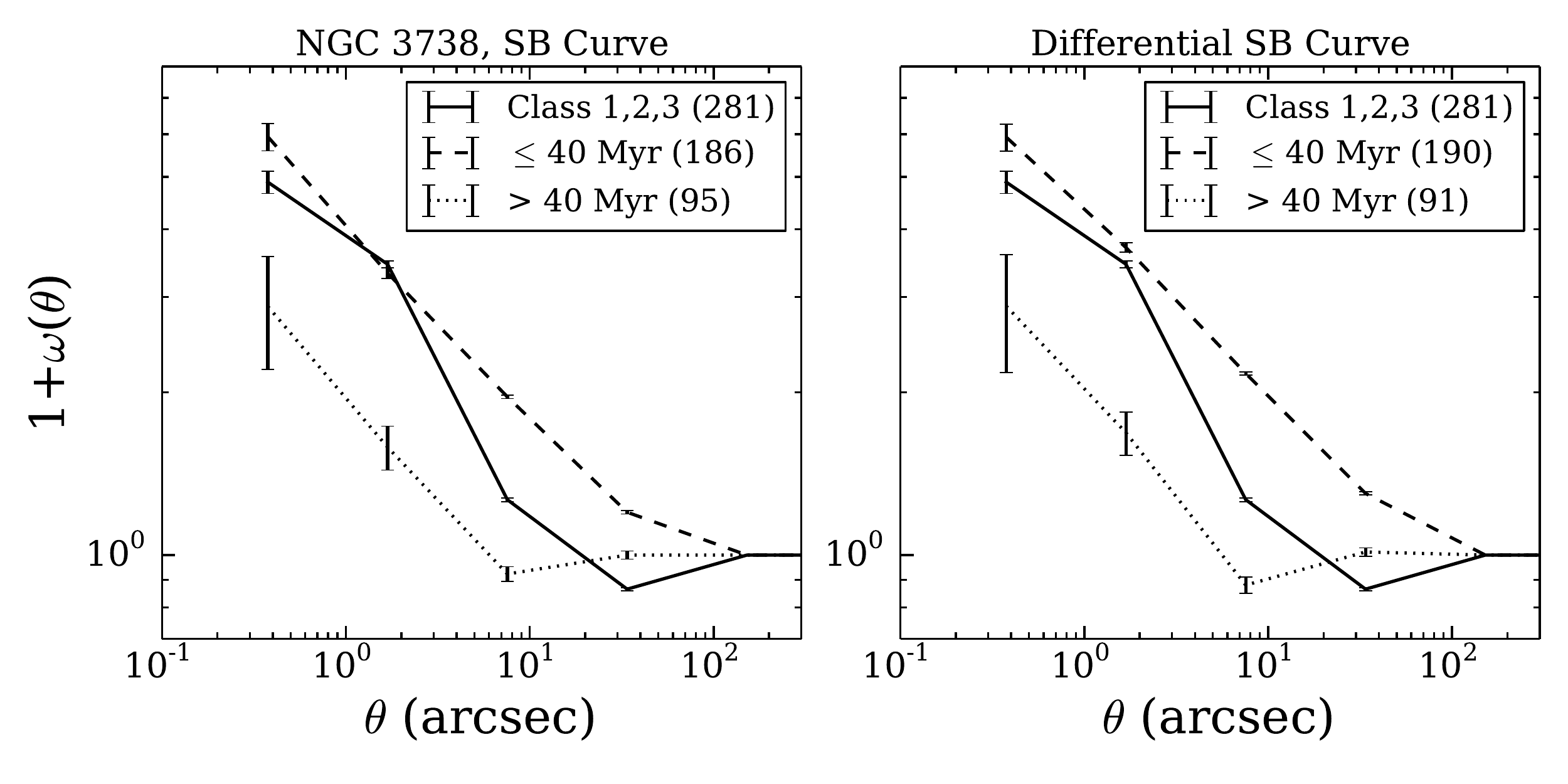}
\caption{
A demonstration on how the different ages impact the two-point correlation function for NGC 3738 for the starburst attenuation law (left; same as Figure \ref{fig:2pcfage}), which is our reference, compared to the differential starburst attenuation law (right) for the three classifications of clusters.  
\label{fig:3738compare}}
\end{figure}

Figure \ref{fig:3738compare} shows the impact that different ages have on the two point correlation function for NGC 3738 when calculated using the starburst attenuation curve (Figure \ref{fig:2pcf}; our reference model), compared to the differential starburst curve (as NGC 3738 has about a quarter of the metallicity of the Milky Way galaxy, we do not consider the Milky Way extinction curve).  The differential starburst law gives a total of four clusters with ages younger than those recovered with the starburst attenuation curve.  The class 1 clusters show the greatest difference in the results, a result of class 1 having the least number of total clusters compared to the total number of class 2 and 3 clusters.  

\begin{figure}
%\epsscale{1.15}
%\plotone{ngc628_2pcf_agecut_4panel_MW_class123.eps}
\includegraphics[scale=.35]{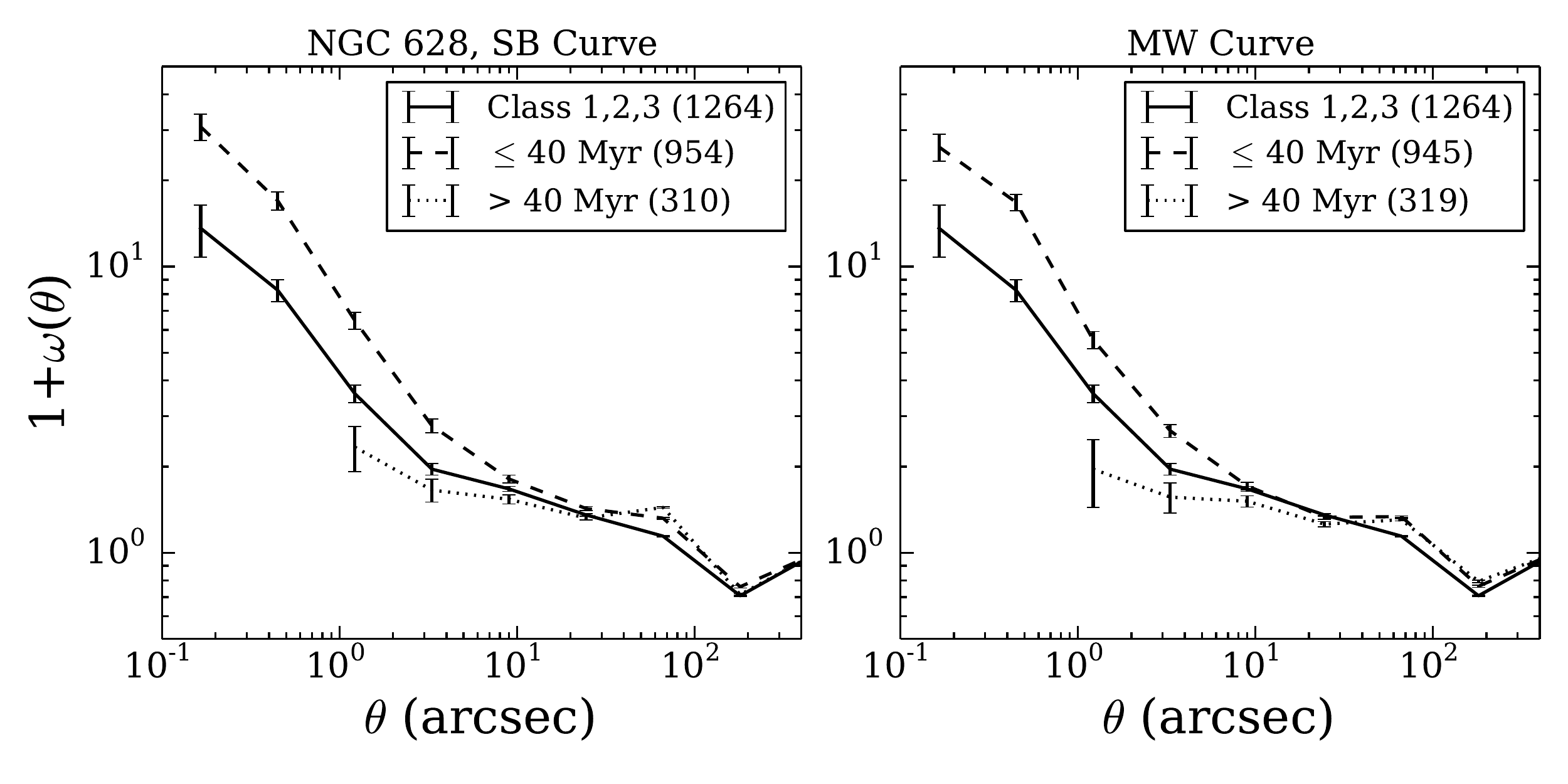}
\caption{
The two-point correlation function for NGC 628, demonstrating how the ages derived for our reference of a starburst attenuation curve (left, same as Figure \ref{fig:2pcfage}) compare to a Milky Way extinction curve (right) for all the classifications of clusters.  The correlation function for the clusters whose properties are determined with a starburst attenuation law versus a Milky Way extinction law does not change.  
\label{fig:628compare}}
\end{figure}

Figure \ref{fig:628compare} shows the two-point correlation function for NGC 628 estimated using the Milky Way extinction curve compared to our reference starburst model.  There are a total of nine clusters that jump between the age bin of lower/higher than 40~Myr when derived with the Milky Way curve.  While there is a large spread in the ages (seen in Figure \ref{fig:compareext} for the Milky Way versus starburst curve, the large number of clusters means the results are fairly insensitive to any particular model.  The only difference is a minor increase in the slope at the lowest spatial bin for the youngest clusters in class 1.  

While the derived ages determined with different models from the SED fits do influence the overall properties of the star clusters, the adopted dust curve does not make a statistical impact on our results nor does it prohibit the comparison of results if the data set is analyzed using two different sets of models.  %The impact increases when there are small numbers, as can be seen for the class 1 clusters in NGC 3738.  

%\bibitem[]{} 


\begin{thebibliography}{}
{\footnotesize
\bibitem[Adamo \etal(2010)]{adamo10} Adamo, A., \"{O}stlin, G., Zackrisson, E., \etal\ 2010, \mnras, 407, 870
\bibitem[Adamo \etal(2017)]{adamo17} Adamo, A., \etal\ 2017, \apj, submitted
\bibitem[Bastian \etal(2007)]{bastian07} Bastian, N., Ercolano, B., Gieles, M., \etal\ 2007, \mnras, 379, 1302
\bibitem[Bastian \etal(2009)]{bastian09} Bastian, N., Gieles, M., Ercolano, B., \& Gutermuth, R. 2009, \mnras, 392, 868
\bibitem[Bate \etal(1998)]{bate98} Bate, M.R., Clarke, C.J., \& McCaughrean, M.J. 1998, \mnras, 297, 1163
\bibitem[Baumgardt \etal(2013)]{baumgardt13} Baumgardt, H., Parmentier, G., Anders, P., \& Grebel, E.K., 2013, \mnras, 430, 676
\bibitem[Beech(1987)]{beech87} Beech, M. 1987, Ap\&SS, 133, 193
\bibitem[Bertin \& Arnouts(1996)]{bertin96} Bertin, E. \& Arnouts, S. 1996, A\&AS, 117, 393
\bibitem[Bhatia \& Hadzidimitriou(1988)]{bhatia88} Bhatia, R.K. Hadzidimitriou, D. 1988, MNRAS, 230, 215
\bibitem[Bhatia(1990)]{bhatia90} Bhatia, R.K., 1990, PASJ, 42, 757
\bibitem[Bothwell \etal(2009)]{bothwell09} Bothwell M.S., Kennicutt R.C., \& Lee J.C. 2009, \mnras, 400, 154
\bibitem[Bressert \etal(2010)]{bressert10} Bressert, E., Bastian, N., Gutermuth, R., \etal\ 2010, \mnras, 409, L54
\bibitem[Calzetti \etal(1989)]{calzetti89} Calzetti, D., Giavalisco, M., \& Ruffini, R. 1989, A\&A, 226, 1
\bibitem[Calzetti \etal(2000)]{calzetti00} Calzetti, D., Armus, L., Bohlin, R.C., \etal\ 2000, \apj, 533, 682
\bibitem[Calzetti \etal(2015)]{calzetti15} Calzetti, D., Lee, J.C., Sabbi, E. \etal\ 2015, \aj, 149, 51
\bibitem[Cardelli \etal(1989)]{cardelli89} Cardelli, J.A., Clayton, G.C., \& Mathis, J.S. 1989, \apj, 345, 245
\bibitem[Combes \etal(2014)]{combes14} Combes, F., Garcia-Burillo, S., Casasola, V., \etal\ 2014, A\&A, 565, 97
\bibitem[Cook \etal(2016)]{cook16} Cook, D.O., Dale, D.A., Lee, J.C., Thilker, D., Calzetti, D., \& Kennicutt, R.C. 2016, \mnras, 462, 3766
\bibitem[da Silva \etal(2012)]{dasilva12} da Silva, R.L., Fumagalli, M., \& Krumholz, M. 2012, \apj, 745, 145
\bibitem[de la Fuente Marcos \& de la Fuente Marcos(2009)]{delafuentemarcos09} de la Fuente Marcos, R. \& de la Fuente Marcos, C. 2009, \apj, 700, 436
\bibitem[de la Fuente Marcos \& de la Fuente Marcos(2010)]{delafuentemarcos10} de la Fuente Marcos, R. \& de la Fuente Marcos, C. 2010, \apj, 719, 104
\bibitem[de Vaucouleurs \etal(1991)]{devaucouleurs91} de Vaucouleurs, G., de Vaucouleurs, A., Corwin, H.G., Jr., \etal\ 1991, Third Reference Catalogue of Bright Galaxies (ver 3.9; New York: Springer)
\bibitem[Dieball \etal(2002)]{dieball02} Dieball, A., M\"{u}llerm H., Grebel, E.K., 2002, A\&A, 391,
547
\bibitem[Efremov \& Elmegreen(1998)]{efremov98} Efremov, Y.N. \& Elmegreen, B.G. 1998, \mnras, 299, 588
\bibitem[Elmegreen \& Falgarone(1996)]{elmegreen96f} Elmegreen, B.G \& Falgarone, E. 1996, \apj, 471, 816
\bibitem[Elmegreen \& Efremov(1996)]{elmegreen96} Elmegreen, B.G. \& Efremov, Y.N. 1996, \apj, 466, 802
\bibitem[Elmegreen \etal(2006)]{elmegreen06} Elmegreen, B.G., Elmegreen, D.M., Chandar, R., \etal\ 2006, \apj, 644, 879
\bibitem[Elmegreen(2008)]{elmegreen08} Elmegreen B.G., 2008, \apj, 672, 1006
\bibitem[Elmegreen \etal(2009)]{elmegreen09} Elmegreen, D.M., Elmegreen B.G., Marcus, M.T., \etal\ 2009, \apj, 701, 306
\bibitem[Elmegreen(2010)]{elmegreen10} Elmegreen, B. G. 2010, Proc. Int. Astronomical Union, IAU Symp. 266, Star Clusters: Basic Galactic Building Blocks Throughout Time and Space 3 (Cambridge: Cambridge Univ. Press)
\bibitem[Elmegreen \etal(2014)]{elmegreen14} Elmegreen, D.M., Elmegreen, B.G., Adamo, A., \etal\ 2014, \apj, 787, L15
\bibitem[Falgarone \etal(1991)]{falgarone91} Falgarone, E., Phillips, T.G., \& Walker, C.K. 1991, \apj, 378, 186
\bibitem[Federrath \etal(2009)]{federrath09} Federrath, C., Klessen, R.S., \& Schmidt, W. 2009, \apj, 692, 364
\bibitem[F\"{o}rster Schreiber \etal(2011)]{forsterschreiber11} F\"{o}rster Schreiber, N.M., Shapley, A.E., Genzel, R., \etal\ 2011, \apj,  739, 45
\bibitem[Gieles \etal(2011)]{gieles11} Gieles, M., Heggie, D. C., \& Zhao, H. 2011, \mnras, 413, 2509
\bibitem[Gieles \& Portegies Zwart(2011)]{gielesportegieszwart11} Gieles, M. \& Portegies Zwart, S.F. 2011, \mnras, 410, L6
\bibitem[Guo \etal(2012)]{guo12} Guo, Y., Giavalisco, M., Ferguson, H.C., \etal\ 2012, \apj, 757, 120
\bibitem[Gouliermis \etal(2014)]{gouliermis14} Gouliermis, D.A., Beerman, L.C., Bianchi, L., \etal\ 2014, in `A Conf. in Honour of David Block and Bruce Elmegreen, Lessons from the Local Group' , ed. K. C. Freeman (New York: Springer) in press (arXiv:1407.0829)
\bibitem[Gouliermis \etal(2014a)]{gouliermis14a} Gouliermis, D.A., Hony, S., \& Klessen, R.S. 2014a, \mnras, 439, 3775
\bibitem[Gouliermis \etal(2015)]{gouliermis15} Gouliermis, D.A., Thilker, D., Elmegreen, B.G., \etal\ 2015, MNRAS, 452, 3508
\bibitem[Grasha \etal(2015)]{grasha15} Grasha, K, Calzetti, D., Adamo, A., \etal\ 2015, \apj, 815, 93
\bibitem[Grasha \etal(2017)]{grasha17} Grasha, K, Elmegreen, B.G., Calzetti, D. \etal\ 2017, \apj, submitted
\bibitem[Immeli \etal(2004)]{immeli04} Immeli, A., Samland, M., Westera, P., \& Gerhard, O. 2004, \apj, 611, 20
\bibitem[Gustafsson \etal(2016)]{gustafsson16} Gustafsson, B, \etal\ 2016, A\&A, astro-ph:1605.02965
\bibitem[Knapen \etal(2006)]{knapen06} Knapen, J.H., Mazzuca, L.M., B\"{o}ker, T., Shlosman, I., Colina, L., Combes, F., Axon, D.J., 2006, A\&A, 448, 489
\bibitem[Krumholz \etal(2015)]{krumholz15} Krumholz, M.R., Adamo, A., Fumagalli, M., \etal\ 2015, \apj, 812, 147
\bibitem[Lada \& Lada(2003)]{lada03} Lada, C.J. \& Lada, E.A. 2003, ARA\&A, 41, 57
\bibitem[Landy \& Szalay(1993)]{landy93} Landy, S.D. \& Szalay, A.S. 1993, \apj, 412, 64
\bibitem[Larson(1981)]{larson81} Larson, R.B. 1981, \mnras, 194, 809
\bibitem[Larson(1995)]{larson95} Larson, R.B. 1995, \mnras, 272, 213
\bibitem[Lee \etal(2009)]{lee09} Lee, J.C., Gil de Paz, A., Tremonti, C., \etal\ 2009, \apj, 706, 599
\bibitem[Leitherer \etal(1999)]{leitherer99} Leitherer, C., Schaerer, D., Goldader, J.D., \etal\ 1999, \apjs, 123, 3
\bibitem[Ma\'{i}z-Apell\'{a}niz(2001)]{maizapellainiz01} Ma\'{i}z-Apell\'{a}niz, J. 2001, \apj, 563, 151
\bibitem[Mandelbrot(1982)]{mandelbrot82} Mandelbrot, B.B. 1982, The Fractal Geometry of Nature (New York: Freeman)
\bibitem[Padoan \etal(2001)]{padoan01} Padoan, P., Kim, S., Goodman, A., Staveley-Smith, L.  2001, \apj, 555, L33
\bibitem[Peebles(1980)]{peebles80} Peebles, P.J.E. 1980, `The Large-Scale Structure of the Universe',  (Princeton, N.J.: Princeton University Press)
\bibitem[Pellerin \etal(2007)]{pellerin07} Pellerin, A., Meyer, M., Harris, J., \& Calzetti, D. 2007, \apj, 658, L87
\bibitem[Pellerin \etal(2012)]{pellerin12} Pellerin, A., Meyer, M., Calzetti, D. \& Harris, J. 2012, AJ, 144, 182
\bibitem[Portegies Zwart \etal(2010)]{portegieszwart10} Portegies Zwart, S.F., McMillan, S.L.W., \& Gieles, M. 2010, ARA\&A, 48, 431
\bibitem[Priyatikanto \etal(2017)]{priyatikanto17} Priyatikanto, R., Kouwenhoven, M.B.N., Arifyanto, M.I., Wulandari, H.R.T., \& Siregar, S. 2017, \mnras, in press
\bibitem[Radburn-Smith \etal(2012)]{radburnsmith12} Radburn-Smith, D.J.,  Ro\v{s}kar, R.  Debattista, V.P., \etal\ 2012, \apj, 753, 138
\bibitem[Ryon \etal(2017)]{ryon17} Ryon, J., \etal\ 2017, \apj, submitted
\bibitem[S\'{a}nchez \etal(2005)]{sanchez05} S\'{a}nchez, N., Alfaro, E. J., \& P\'{e}rez, E. 2005, \apj, 625, 849
\bibitem[S\'{a}nchez \& Alfaro(2008)]{sanchez08} S\'{a}nchez, N. \& Alfaro, E.J. 2008, \apjs, 178, 1
\bibitem[S\'{a}nchez \& Alfaro(2009)]{sanchez09} S\'{a}nchez, N. \& Alfaro, E.J. 2009, \apj, 696, 2086
\bibitem[Scalo(1985)]{scalo85} Scalo, J.M. 1985 in Protostars and Planets II, ed. D.C. Black \& M.S. Mathews (Tucson: Univ. of Arizona Press), 201
\bibitem[Scheepmaker \etal(2009)]{scheepmaker09} Scheepmaker, R.A., Lamers, H.J.G.L.M., Anders, P., \& Larsen, S.S. 2009, \aap, 494, 81
\bibitem[Schlafly \& Finkbeiner (2011)]{schlafly11} Schlafly, E.F. \& Finkbeiner, D.P. 2011, \apj, 737, 103
\bibitem[Smaji\'{c} \etal(2015)]{smajic15} Smaji\'{c}, S., Moser, L., Eckart, A., \etal\ 2015, A\&A, 583, A104
\bibitem[Subramaniam \etal(1995)]{subramaniam95} Subramaniam, A., Gorti, U., Sagar, R., Bhatt, H. 1995, A\&A, 302, 86
\bibitem[Verdes-Montenegro \etal(2000)]{verdes-montenegro00} Verdes-Montenegro, L., Bosma, A., \& Athanassoula, E. 2000, A\&A, 356, 827
\bibitem[Whitmore \etal(2005)]{whitmore05} Whitmore, B.C, Gilmore, D., Leitherer, C. \etal\ 2005, \aj, 130, 2104
\bibitem[Wright \etal(2014)]{wright14} Wright, N.J., Parker, R.J., Goodwin, S.P., \& Drake, J.J. 2014 \mnras 438, 639
\bibitem[Zackrisson \etal(2011)]{zackrisson11} Zackrisson, E., Rydberg, C.-E., Schaerer, D., \"{O}stlin, G., \& Tuli, M. 2011, \apj, 740, 13
\bibitem[Zhang \etal(2001)]{zhang01} Zhang, Q., Fall, S.M., \& Whitmore, B.C. 2001, \apj, 561, 727
}
\end{thebibliography}
\end{document}